\documentclass[prd,twocolumn,preprintnumbers,nofootinbib]{revtex4-1}

\usepackage[utf8]{inputenc}

\usepackage{amsmath}
\usepackage{epsfig}
\usepackage{graphicx}
\usepackage{graphics}
\usepackage{color}
\usepackage{aas_macros}
\usepackage{bbm}
\usepackage[normalem]{ulem}
\usepackage[breaklinks, plainpages=false, colorlinks=true, anchorcolor=cyan, linkcolor=red, citecolor=cyan, urlcolor=magenta, bookmarks=false]{hyperref}

\usepackage[caption=false]{subfig}
\usepackage[title]{appendix}
\usepackage{tikz,tikz-3dplot}

\begin{document}
\renewcommand{\thefigure}{\arabic{figure}}
\setcounter{figure}{0}

 \def\I{{\rm i}}
 \def\E{{\rm e}}
 \def\D{{\rm d}}

\bibliographystyle{apsrev}

\title{Detecting hierarchical stellar systems with LISA}

\author{Travis Robson}
\affiliation{eXtreme Gravity Institute, Department of Physics, Montana State University, Bozeman, Montana 59717, USA}

\author{Neil J. Cornish}
\affiliation{eXtreme Gravity Institute, Department of Physics, Montana State University, Bozeman, Montana 59717, USA}

\author{Nicola Tamanini}
\affiliation{Max-Planck-Institut f\"ur Gravitationsphysik, Albert-Einstein-Institut, Am M\"uhlenberg 1,
14476 Potsdam-Golm, Germany}

\author{Silvia Toonen}
\affiliation{Anton Pannekoek Institute for Astronomy, University of Amsterdam, NL-1090 GE Amsterdam, The Netherlands}

\begin{abstract} 
A significant fraction of stars are members of gravitationally bound hierarchies containing three or more components. Almost all low mass stars in binaries with periods shorter three days are part of a hierarchical system. We therefore anticipate that a large fraction of compact galactic binaries detected by the Laser Interferometer Space Antenna (LISA) will be members of hierarchical triple or quadruple system. The acceleration imparted by the hierarchical companions can be detected in the gravitational wave signal for outer periods as large as 100 years. For systems with periods that are shorter than, or comparable to, the mission lifetime, it will be possible to measure the period and eccentricity of the outer orbit.  LISA observations of hierarchical stellar systems will provide insight into stellar evolution, including the role that Kozai-Lidov oscillations play in driving systems towards merger.
\end{abstract}

\maketitle

\section{Introduction}

The Laser Interferometer Space Antenna (LISA)~\cite{Audley:2017drz} is expected to individually resolve the signals from tens of thousands of compact galactic binaries during its nominal four year mission lifetime~\cite{Cornish:2017vip}. Roughly 13\% of low mass stellar systems contain three or more stars~\cite{Rag10, 2014AJ....147...87T, Fuh17}, and roughly $96$\% of low mass binaries with periods shorter than 3 days are part of a larger hierarchy~\cite{2006A&A...450..681T,2006AJ....131.2986P}. While the multiplicity distribution for ultra-compact binaries  is currently unknown, it is reasonable to expect that a significant fraction of compact galactic binary systems detected by LISA will be members of a hierarchical system. Indeed, dynamical effects in hierarchical systems such as eccentric Kozai-Lidov oscillations can cause the inner binary orbit to harden, potentially enhancing the fraction of compact binary systems with companions~\cite{2011ApJ...741...82T,Fang:2017wav, Toonen:2017yct, Fab07}. 

The hierarchal companions to the ultra-compact binaries detected by LISA will impart accelerations on the center of mass of the binary that can lead to observable doppler shifts in the signals. This effect has previously been considered in the context of LISA observations of extreme mass ratio binaries~\cite{Yunes:2010sm}, and merging black hole binaries detected by LIGO and LISA~\cite{Meiron:2016ipr,Bonvin:2016qxr,Inayoshi:2017hgw,Randall:2018lnh}. The mathematical description is essentially identical to that in pulsar timing, where the orbital parameters of pulsars found in binary systems can be inferred from modulations of the radio pulses~\cite{Edwards:2006zg}. One difference between the radio and gravitational wave analyses is that wavelengths of the gravitational waves are significantly larger than the gravitational radii of the stars, which modifies the calculation of the Shapiro time-delay. 

Here we consider LISA observations of compact galactic binaries in hierarchical systems and identify three main regimes that are governed by the ratio of the outer orbital period to the observation time: (1) When the outer period is much larger than the observation time the hierarchical orbit imparts an overall unobservable doppler shift (2) When the outer period is up to a factor ten to so larger than the observation time the influence of the companion can be detected (3) When the outer period is shorter than or comparable to the observation time, the eccentricity and period of the hierarchical orbit can be inferred. In rare cases a fourth regime can occur where (4) the acceleration due to the hierarchal perturber can be mistaken for frequency changes due to gravitational wave emission or mass transfer. This regime occurs when the outer period is larger than the observation time, and the chirp mass and gravitational wave frequency of the compact binary lie in a narrow range of values. The precise location of the boundaries between the four cases depends on several factors, including the signal-to-noise ratio, the gravitational wave frequency, the mass ratio between the inner binary and the perturber, and the eccentricity of the outer orbit. Using a simple Fisher matrix based estimate for when the frequency change of a nearly monochromatic signal can be detected, we arrive at the condition that, on average, the outer binary can be detected when the period of the outer orbit, $P_2$ obeys the inequality
\begin{eqnarray}
&& P_{2} \lesssim 43.2 \text{ yrs}\left(\frac{\rho}{10} \cdot \frac{m_{c}}{1.0 M_{\odot}}\cdot \frac{f}{5 \text{ mHz}}\right)^{3/4}\left(\frac{m_{2}}{2 M_{\odot}}\right)^{-1/2} \nonumber \\
&&  \quad \quad \times \left(\frac{T_{\mbox{\tiny obs}}}{4 \text{ yr}}\right)^{3/8} \left(\frac{1+\frac{1}{2}e_{2}^{2}}{(1-e_{2}^{2})^{5/2}}\right)^{3/8}\, .
\label{eq:fdot_meas}
\end{eqnarray}
This expression is valid for $P_{2} > T_{\mbox{\tiny obs}}$, where $T_{\mbox{\tiny obs}}$ is the observation time. For shorter periods higher derivatives of the frequency change with respect to time need to be accounted for. Other quantities that appear in the expression are the signal-to-noise ratio $\rho$, the mass of the perturber $m_c$, the gravitational wave frequency $f$, the total mass of the system $m_2$ and the eccentricity of the hierarchical orbit $e_2$, and we work in geometrical units with $G=c=1$. To derive this expression we computed the root-mean-square (RMS) line-of-sight acceleration of inner binary due to the distant companion, averaging over the orbital period and orientation. Note that some systems will be detectable with longer periods if the orientation and phase of the orbit is more favorable. Also note that LISA is expected to detect hundreds of galactic binaries with signal-to-noise ratios (SNR) $\rho > 100$, and for these systems it will be possible to detect systems with $P_2 > 100$ years.


The outline of the paper is as follows: In \S\ref{astro} we review what is known about compact binaries in hierarchical systems. In \S\ref{models} we summarize the models and methods used in our study. The orbital model is described in more detail in \S\ref{orbit} and the gravitational wave modeling is outlined in \S\ref{gw}. The detectability of hierarchical companions is considered in \S\ref{detect}, and the characterization of the orbits is investigated in \S\ref{characterize}. The possibility of confusing the acceleration caused by a distant perturbed with orbital evolution due to radiation reaction or mass transfer is discussed in \S\ref{ambiguous}. We conclude with a summary and discussion of future studies in  \S\ref{discus}.

\section{Compact Binaries In Hierarchical Systems}\label{astro}

The majority of stars are members of multiple systems, including binaries, triples and higher-order hierarchies.
The triple fraction is best known for stellar systems with main-sequence components, 
in particular for lower mass stars of F- and G-type. In the 25pc all-sky survey of \cite{Rag10}, a multiple fraction of 11\% was found, whereas recent updates to the sample advocate fractions up to 20\%~\cite{Fuh17}. From a larger sample of F- and G-type stars up to 67pc, \cite{Tok14b} finds a multiple fraction of 13\%. 
The triple fraction increases for higher mass stars, with a lower limit of 30\% for O- and B-type stars \citep{San14}, and 44\% in a sample of 18 Cepheids \citep{Eva05} (not corrected for biases). 

The period distribution of the inner and outer orbits of triples with F- and G-type primaries are distributed similarly as those of binaries, however, with the additional constraint that the triple is dynamically stable \cite{Tok14a}. As a result the inner orbits tend to be more compact, leading more often to mass transfer episodes and compact binaries (Toonen et al. in prep.).  
Besides the initial structure of the triple, three-body dynamics can provide additional means to harden the inner binary. 
The classical low-order approximation of the three-body problem are the Lidov-Kozai cycles in which the mutual inclination between the two orbits and the eccentricity of the inner binary vary periodically \cite{Lid62,Koz62}. For a comprehensive
review of the Lidov-Kozai effect see \cite{Nao16}, and for the evolution of stellar triples \cite{Too16}. 
Due to the high eccentricities achieved during the Lidov-Kozai cycles, 
the Lidov-Kozai mechanism is linked to a  variety of exotic astrophysical phenomena,
such as stellar mergers \citep{Tho11, Per12, Ham13}, X-ray binaries \citep{Iva10}, blue stragglers \citep{Per09, Nao14}
as well as enhanced dissipation through gravitational wave emission and tides \citep{Kis98,Maz79}. 
Due the latter mechanism, also known as high-eccentricity migration or Lidov-Kozai cycles with tidal friction (LKCTF), the inner binary tightens forming hot Jupiters \citep[e.g.][]{Cor11, Pet15, Wu03} and compact binaries \citep{Kis98, Fab07, Maz79}; 
Observationally roughly 96\% of low-mass binaries with periods shorter than three days have outer companions \citep{2006A&A...450..681T,2006AJ....131.2986P}.

In the context of GW sources, Lidov-Kozai cycles are relevant, as the gravitational wave inspiral time of a close (inner) binary with compact objects can be significantly reduced, if an outer star is present.
Whereas isolated compact binaries need to be formed at periods $\lesssim0.3$ days to merge within a Hubble time, the presence of an outer companion extends the inner period range to hundreds of days if LKCTF is efficient. 
Even wider inner orbits can be brought to merge or collide if the triple system is weakly hierarchical for which the secular perturbation theory breaks down \citep{Ant12, Kat12, Ant17, Too17}. Such mergers of compact objects occur in orbits with higher residual eccentricities  \citep[e.g.][]{Gul06,Set13,Ran18}.

On the observational side, our knowledge of the triple fraction and orbital structures of triples with compact objects is limited. 
The highly complete sample of WDs within 20pc from the Sun, contains one to two triples with an inner compact double WD, showing that indeed its possible to form such object \citep[e.g.][]{Too17}. Moreover, out of about 130 objects in total, there is only one confirmed isolated compact double WD and four candidates, indicating that triple sources are relatively abundant.

When shifting our attention from compact double white dwarfs to wide systems, there are only two such binaries within 20pc. 
This is in contradiction to theory, from which one would expect 15-30 such systems within 20 pc \citep{Too17}. As destruction mechanism (e.g. dynamical interactions or stellar winds) are not efficient enough to explain the discrepancy, it has been claimed that the progenitor systems are not formed as efficiently as expected \citep[e.g.][]{Too17} or that the wide double white dwarfs have been missed observationally \citep{Kle17}, however in the state-of-the-art sample of Gaia no new wide double white dwarfs were found within 20 pc (Hollands et al. in prep.).

Interesting to mention is PSR J0337+1715, the millisecond pulsar in a hierarchical triple with two white dwarfs \citep{Ran14} with periods of 1.6d and 327d. As both white dwarfs are low-mass helium dwarfs, the system demonstrates that it is possible in nature for a triple to survive several phases of mass transfer \citep[see e.g.][for possible formation scenarios]{Tau14, Sab15}, and have outer periods in the range of the LISA mission lifetime.

\section{Summary of Models and Methods}\label{models}
The natural separation of scales found in hierarchical systems allows us to make a number of simplifying assumptions. The few-body Hamiltonian for a hierarchical system can be expanded in the ratio of the semi-major axes yielding terms at monopole, quadrupole, octapole and higher orders~\cite{2000ApJ...535..385F}. Here we are mostly interested in 2:1 and 2:2 component hierarchies where the semi-major axis of the binary components are much smaller than semi-major axis of the overall system. Because the hierarchical periods we are considering will be comparable to or larger than the observation time, we can restrict our analysis to the leading order, Newtonian monopole interactions. In this approximation, the motion of the binaries separates from that of the hierarchical system, and each can be treated as a separate Keplarian system. The center of mass of the inner binary follows a Keplarian orbit around the distant perturber. We are justified in doing this since the Kozai-Lidov~\cite{1962AJ.....67..591K, 1962P&SS....9..719L} and eccentric Kozai-Lidov~\cite{2000ApJ...535..385F, 0004-637X-793-2-137} oscillations induced by the quadrupole and octapole terms occur on timescales that are long compared to the period of the hierarchical orbit, and very much longer then the observation time. The same is true for the high-order post-Newtonian effects such as periastron precession.

We allow for the outer hierarchical orbit to be eccentric, but make the simplifying assumption that the inner orbit responsible for the gravitational wave emission is circular. We can justify this choice in two ways. First, gravitational radiation acts to quickly circularize orbits, and second, even if effects such as Kozai-Lidov oscillations have managed to maintain the eccentricity of the inner binary, our results will be little changed, at least for moderate eccentricities. The reasoning is as follows: for slowly evolving, moderately eccentric systems the gravitational wave signal can be expressed as a sum of circular binaries with periods at harmonics of the orbital period. The separation of these harmonics in frequency is very much larger than the sidebands imparted by the hierarchical orbit, so there is zero confusion between the two effects. The sum of circular binary signals for an eccentric system contains almost identical information to that of a single circular binary for the purposes of the current analysis, so in the interests of computational efficiency we neglect the eccentricity of the inner binary.

To assess the detectability of the distant companion and the accuracy with which the parameters of the orbits can be inferred we use a mixture of methods. To make quick estimates and derive analytic scalings we compute Fisher information matrices, and to spot check these estimates and provide more detailed results we employ Bayesian inference via the Markov Chain Monte Carlo algorithm.

\section{Hierarchical Orbit Model}\label{orbit}


In this section we derive how the perturbing companion affects the center-of-mass motion of the inner binary in the hierarchical orbit which will impart perturbations to the gravitational waveform. We desire to extract the line-of-sight component of the inner binary’s center-of-mass velocity. For an isolated binary, its center-of-mass is stationary with respect to the solar system barycenter (ignoring unobservable constant peculiar velocities), but this line-of-sight component of the induced center-of-mass will create a time-dependent red-shift as seen in the barycenter frame. We will use ``1'' subscripts to denote orbital parameters of the inner gravitational wave emitting binary comprised of masses $m_{a}$ and $m_{b}$ for a total mass of $m_{1}$. The subscript ``2'' will denote the Keplerian outer orbit describing the motion of the perturber $m_{c}$ and the monopole mass of the inner binary. In our hierarchical approximation, in which we essentially have a circular Keplerian orbit emitting gravitational wave visible to LISA inside of a larger outer Keplerian orbit that is governed by 

\begin{equation}
\textbf{a}_{2} = -\frac{G m_{2}}{r_{2}^{2}}\hat{\textbf{r}}_{2} \,\, ,
\end{equation}
where $\textbf{a}_{2}$ is the relative acceleration and $\hat{\textbf{r}}_{2} = \hat{\textbf{r}}_{c} - \hat{\textbf{r}}_{1}$ is the unit separation vector as defined in an inertial coordinate system of the triple, and $m_{2} = m_{a} + m_{b} + m_{c}$. 

The solution for the orbital motion is then 

\begin{equation}
\textbf{r}_{2}(t) = r_{2}(t) \left( \cos\varphi_{2}, \sin\varphi_{2},0 \right) \, \, ,
\label{eq:outer_separation_vector}
\end{equation}
where 

\begin{equation}
r_{2}(t) = \frac{a_{2}(1-e_{2}^{2})}{1+e_{2}\cos\varphi_{2}} \, \, ,
\end{equation}
defining the standard Keplerian ellipse. The quantities introduced are defined as follows: $\varphi_{2}$ is the orbital phase of the outer orbit, $e_{2}$ and $a_{2}$ are its eccentricity and semi-major axis respectively. In order to relate the orbital phase to time, its convention to introduce the eccentric anomaly which is a middle man angle related to the orbital phase by the geometric relation

\begin{equation}
\varphi = \varphi_{0} + 2 \tan^{-1}\left(\sqrt{\frac{1+e}{1-e}}\tan\frac{u}{2}\right) \, \, .
\end{equation}
where $\beta_{2} = (1-\sqrt{1-e_{2}^{2}})/e_{2}$. 
The eccentric anomaly is then related to time through Kepler's equation 

\begin{equation}
n_{2}(t-T_{2}) = u_{2} - e_{2}\sin u_{2} \,\, ,
\end{equation}
where $n_{2} = \sqrt{m_{2}/a_{2}^{3}}$ defines the mean motion, or mean angular frequency associated with an orbit. The mean motion is related to $P_{2}$, the outer period, $n_{2} = 2\pi/P_{2}$, and lastly, the parameter $T$ is the time of pericenter passage, a constant of integration.  

The desired velocity of the inner binary's center-of-mass is simply obtained by $\textbf{v}_{1} = (m_{c}/m_{2})\textbf{v}_{2}$, and differentiating equation (\ref{eq:outer_separation_vector}) 

\begin{equation}
\textbf{v}_{1} = \frac{m_{c}}{m_{2}}\sqrt{\frac{G m_{2}}{p_{2}}}\left(-\sin\varphi_{2}, \cos\varphi_{2} + e_{2},0 \right) \,\, ,
\label{eq:v1_vector}
\end{equation}
where $p_2 = a_2 (1-e_2^2)$ is the semi-latus rectum.
Up to this moment we have been working in a coordinate system where the outer orbit defines the xy-plane. We must rotate our system to properly orient it into the coordinate system used by our detector model: the solar system barycenter frame. This may be accomplished through a series of Euler rotations: a rotation of $-\omega_{2}$, around the barycenter's z-axis, then by $-\iota_{2}$ around the new x-axis, and finally $-\Omega_{2}$ around the new z-axis, which are given by the following matrices:

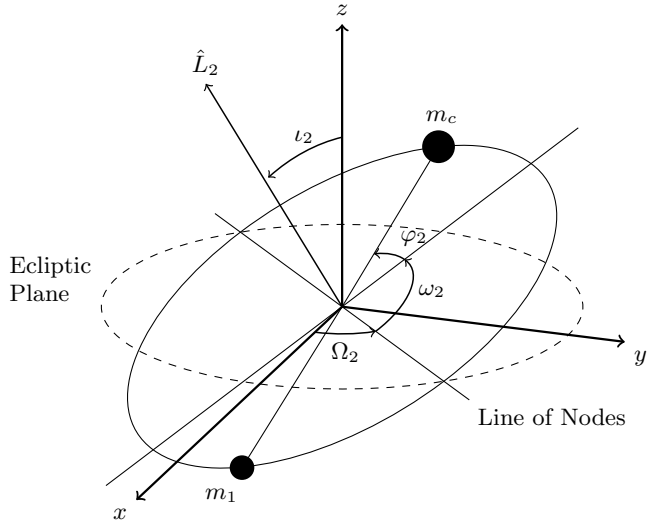
\begin{figure}[htp]
	\begin{centering}
		\tdplotsetmaincoords{70}{110}
		\begin{tikzpicture}[tdplot_main_coords,scale=4]
		\pgfmathsetmacro{\r}{.8}
		\pgfmathsetmacro{\O}{45} 
		\pgfmathsetmacro{\i}{30} 
		\pgfmathsetmacro{\f}{35} 
		
		\coordinate (O) at (0,0,0);
		
		\draw [line width=0.3mm, ->] (O) -- (2,0,0) node[anchor=north east] {$x$};
		\draw [line width=0.3mm, ->] (O) -- (0,1,0) node[anchor=north west] {$y$};
		\draw [line width=0.3mm, ->] (O) -- (0,0,1) node[anchor=south] {$z$};
		
		\node at (0,-\r,0) [left,text width=4em] {Ecliptic Plane};
		
		\tdplotdrawarc[dashed]{(O)}{\r}{0}{360}{}{}
		
		\tdplotsetrotatedcoords{\O}{0}{0}
		
		\draw [tdplot_rotated_coords] (-1,0,0) -- (1,0,0) node [below right] {Line of Nodes};
		\tdplotdrawarc[line width=0.2mm, ->]{(O)}{.33*\r}{0}{\O}{anchor=north}{$\Omega_{2}$}
		
		\tdplotsetrotatedcoords{-\O}{\i}{0}
		\tdplotdrawarc[tdplot_rotated_coords]{(O)}{\r}{0}{360}{}{}  
		\begin{scope}[tdplot_rotated_coords]
		\draw[line width=0.2mm, ->] (O) -- (0,0,1) node [above] {$\hat{L}_{2}$};
		\draw (1,0,0) -- (-1,0,0);
		\tdplotdrawarc[line width=0.2mm, ->]{(O)}{.33*\r}{90}{180}{anchor=west}{$\omega_{2}$}
		\coordinate (P) at (180+\f:\r);
		\draw (O) -- (P);
		\tdplotdrawarc[line width=0.2mm, ->]{(O)}{.33*\r}{180}{180+\f}{anchor=south west}{$\varphi_{2}$}
		\end{scope}
		
		\tdplotsetrotatedcoords{-\O+\f}{\i}{0}
		\tdplotsetrotatedcoordsorigin{(P)}
		\begin{scope}[tdplot_rotated_coords,scale=.2,thick]
		\fill (P) circle (2.0ex) node [shift={(-.5,-0.2)}]{$m_{c}$};
		\end{scope}
		
		\coordinate (Q) at (8:2.*\r);
		\draw (O) -- (Q);
		\tdplotsetrotatedcoords{-(-\O+\f)}{\i}{0}
		\tdplotsetrotatedcoordsorigin{(Q)}
		\begin{scope}[tdplot_rotated_coords,scale=.2,thick]
		\fill (Q) circle (1.5ex) node [shift={(.5,-0.2)}] {$m_{1}$}; 
		\end{scope}
		\tdplotsetthetaplanecoords{-\f}
		\tdplotdrawarc[line width=0.2mm, tdplot_rotated_coords,->]{(O)}{.75*\r}{0}{\i}{anchor=south}{$\iota_{2}$} 
		\end{tikzpicture}
		\caption{\label{fig:tripe_geo} The geometry of the outer orbit consisting of $m_{c}$ and the inner binary's monopole moment $m_{1}$ as displayed above, where the orientation angles are with respect to the solar system barycenter frame where the xy-plane define the Ecliptic plane.}
	\end{centering}
\end{figure}

\begin{align}
\mathbbm{R}_{1} &= \left( \begin{array}{c c c}
\cos\omega_2 & -\sin\omega_2 & 0 \\
\sin\omega_2 & \cos\omega_2 &0 \\
0 & 0 & 1
\end{array} \right) \, \, ,\\
\mathbbm{R}_{2} &= \left( \begin{array}{c c c}
1 & 0 & 0 \\
0 & \cos\iota_2 &-\sin\iota_2 \\
0 & \sin\iota_2 & \cos\iota_2
\end{array} \right) \, \, ,\\
\mathbbm{R}_{3} &= \left( \begin{array}{c c c}
\cos\Omega_2 & -\sin\Omega_2 & 0 \\
\sin\Omega_2 & \cos\Omega_2 &0 \\
0 & 0 & 1
\end{array} \right) \, \, ,
\end{align}
operated in the order $\mathbbm{R} = \mathbbm{R}_{3}\cdot\mathbbm{R}_{2}\cdot\mathbbm{R}_{1}$. As shown in figure \ref{fig:tripe_geo} the line of ascending nodes (labelled in the figure) is defined by a rotation of angle $\Omega_{2}$ from the barycenter x-axis to where the outer orbital plane intersects the Ecliptic. The angle $\omega_{2}$ defines the rotation angle from the line of nodes to the argument of periapsis (whose position is given by the solid line passing through the semi-major axis of the orbit), and $\iota_{2}$ is the inclination angle, i.e. the angle between the outer orbit's angular momentum $\hat{L}_{2}$ (of course neglecting any contribution to angular momentum due to the fact that the inner binary is extended and has an orbit of its own) and the z-axis of the barycenter coordinates.

Finally, we may construct the desired quantity: the line-of-sight velocity $v_{\mbox{\tiny$\parallel$}}$. We can use the line-of-sight vector $\hat{\textbf{n}}$ pointing from the origin of the barycenter coordinates to the triple center-of-mass. Due to the large distances involved we will assume that the sky location $(\theta,\phi)$ of the inner binary and of the triple's center-of-mass are located at the same point on the sky. This vector is of course given by $\hat{\textbf{n}} = \left(\sin\theta\cos\phi, \sin\theta\sin\phi,\cos\theta\right)$. At last we obtain the expression

\begin{align}
v_{\mbox{\tiny$\parallel$}}(t) =& \hat{\textbf{n}}\cdot\mathbbm{R}\cdot\textbf{v}_{1} \,\,,\\
=& \hat{\textbf{n}}\cdot\mathbbm{R}\cdot\left(\frac{m_{c}}{m_{2}}\textbf{v}_{2}\right) \, \, , \nonumber \\
=&~ \frac{m_{c}}{m_{2}}\sqrt{\frac{G m_{2}}{p_{2}}} \nonumber\\
&\times \bigg\lbrace C(\theta,\iota_{2},\phi-\Omega_{2}) \left[\cos(\varphi_{2}+\omega_{2}) + e_{2} \cos\omega_{2} \right] \nonumber \\
&- S(\theta,\phi-\Omega_{2})\left[\sin(\varphi_{2}+\omega_{2}) + e_{2} \sin\omega_{2} \right]\bigg\rbrace  \,\, ,
\label{eq:v_los}
\end{align}
where $C(\theta,\iota_{2},\phi-\Omega_{2}) = \cos\theta\sin\iota_{2} + \sin\theta\cos\iota_{2}\sin(\phi-\Omega_{2})$ and $S(\theta,\phi-\Omega_{2})=\sin\theta\cos(\phi-\Omega_{2})$. 
In the above form it is unclear how many extra parameters are truly involved in the modeling of the triple system, so we re-write the line-of-sight velocity in the simpler form
\begin{equation}
v_{\mbox{\tiny$\parallel$}}(t) = \mathcal{A}_{2} \left[\sin(\varphi_{2}+\varpi) + e_{2} \sin(\varpi) \right] \,\, ,
\label{eq:v_LOS}
\end{equation}
where $\mathcal{A}_{2} = \frac{m_{c}}{m_{2}}\sqrt{\frac{m_{2}}{p_{2}}} \bar{A}$ and $\bar{A}^{2} = C^{2} + S^{2}$ and finally $\varpi = \omega + \bar{\phi}$ where $\tan\bar{\phi} = \frac{C}{-S}$. This form the of the line of sight velocity allows us to see what parameter we may hope to extract by using this model for the triple. 
To specify $v_{\mbox{\tiny$\parallel$}}(t)$ we needed the parameters $n_{2}, e_{2}, T_{2}, \iota_{2}, \omega_{2}, \Omega_{2}, m_{c}, m_{2}$ (note that the sky location angles are part of the binary model), but unfortunately we do not have access to all of these parameters due to degeneracies in the model which can be seen from equation (\ref{eq:v_LOS}). 
The parameters $\omega_{2}$, $\Omega_{2}$, and $\iota_{2}$ get lumped into $\mathcal{A}_{2}$, and $\varpi$, leaving us in no position to parse the dynamically interesting $\iota_{2}$ from other orientation angles. This amplitude has an average value of roughly $0.77$ which will be used in the analysis contained in later sections. Also hidden in $\mathcal{A}_{2}$ are $m_{2}$, and $m_{c}$ which we will not have access to individually. The orbital phase $\varphi_{2}$  contains the uninteresting parameter $T_{2}$ and is also controlled strongly by the mean motion $n_{2}$ and eccentricity $e_{2}$ parameters. We are now in a position to incorporate the line-of-sight velocity into the gravitational waveform. For eclipsing systems the Shapiro time delay can break some of the degenerates and allow us to measure $\iota_{2}$. We will leave the analysis of the gravitational wave Shapiro time delay to future work.

\section{Gravitational Wave and Instrument Model}\label{gw}



We will first briefly review the gravitational wave model for an isolated galactic binary as seen by LISA, and then incorporate the affects due to the companion body. The plus and cross gravitational wave polarizations in the compact binary's barycenter frame are given by

\begin{align}
h_{+} &= \frac{2 \mathcal{M}}{D_{\mathrm{L}}} \left(\pi f_{\mathrm{gw}}(t)\right)^{2/3}\left(1+\cos^{2}\iota_{1}\right) \cos \Psi_{\mathrm{gw}} \,\,,\\
h_{\times} &=- \frac{4 \mathcal{M}}{D_{\mathrm{L}}} \left(\pi f_{\mathrm{gw}}(t)\right)^{2/3}\cos\iota_{1} \sin \Psi_{\mathrm{gw}} \,\,,\\
\end{align}
where $D_{\mathrm{L}}$ is the luminosity distance, $\mathcal{M} = (m_{a}m_{b})^{3/5}/m_{1}^{1/5}$ is the chirp mass, $f_{\mathrm{gw}}$ is the instantaneous gravitational wave frequency (as measured in the compact binary's barycenter frame), $\Psi_{\mathrm{gw}}$ the corresponding gravitational wave phase, and lastly $\iota_{1}$ is the inclination of the inner binary i.e. $\cos\iota_{1} = \hat{L}_{1}\cdot\hat{z}$. One may obtain the gravitational wave phase from the frequency through $\Psi_{\mathrm{gw}} = 2\pi \int^{t} f_{\mathrm{gw}}(t')dt' + \phi_{0}$ where $\phi_{0}$ is an arbitrary phase shift.  

For galactic binaries whose orbital evolution is dominated by gravitational wave radiation reaction, the frequency evolution is given by

\begin{equation}
f_{\mathrm{gw}}(t) = \frac{1}{8\pi \mathcal{M}}\left(\frac{5 \mathcal{M}}{t_c - t}\right)^{3/8} \,\, ,
\end{equation}
where $t_{c}$ is the time of coalescence for the binary; a $3$ mHz, $0.265 ~M_{\odot}$ galactic binary will merge in $1$ million years. The number of $1/T_{\rm obs}$ frequency bins a fiduciary source evolves through over the LISA mission lifetime is given by \cite{Seto02}
\begin{align}
\dot{f} T_{\mathrm{obs}}^{2} = &~5.1 \left(\frac{\mathcal{M}}{0.32 M_{\odot}}\right)^{5/3} \left(\frac{f}{5 \mathrm{ mHz}}\right)^{11/3}\left(\frac{T_{\mathrm{obs}}}{4 \mathrm{ yrs}}\right)^{2} \label{eq:fdot_RR}\\
\ddot{f} T_{\mathrm{obs}}^{3} = &~1.5\times 10^{-4} \nonumber \\
&\times \left(\frac{\mathcal{M}}{0.32 M_{\odot}}\right)^{10/3} \left(\frac{f}{5 \mathrm{ mHz}}\right)^{19/3}\left(\frac{T_{\mathrm{obs}}}{4 \mathrm{ yrs}}\right)^{3} \,\, .
\end{align}
The strong frequency dependence in these expression implies that the higher frequency sources will have more measurable chirps. It is this frequency dependence that will allow us to determine the physics responsible for the evolution of a population of binaries. A similar order of magnitude frequency evolution is experienced by galactic binaries which involve stable mass transfer \cite{Marsh03}.  A key difference is that mass transfer tends to widen orbits leading to a frequency decrease over time.
The mild evolution in gravitational wave frequency lends itself to a Taylor expansion:
\begin{equation}
f_{\mathrm{gw}}  = f + \dot{f} t + \frac{1}{2}\ddot{f}t^{2} \,\,,
\label{eq:f_src}
\end{equation}
whose coefficients are determined by the dynamics at play in the binary. We shall refer to $f$ (and the equivalently red-shifted version during the triples discussion) as the carrier frequency.

Cornish \& Littenberg~\cite{CornLitt07} present a frequency domain model $\tilde{h}$ for galactic binaries measured by LISA. Under the rigid adiabatic approximation to the LISA motion one is able to perform a fast-slow decomposition, due to the slowly evolving amplitude (varying on timescales of a year mostly due to LISA's motion) and the fast varying phase due the larger carrier frequency (corresponding to orbital periods of minutes to hours for galactic binaries) of the waveform allowing a rapid evaluation of the waveform.

\begin{figure}[htp]
	\begin{centering}
		\includegraphics[clip=true,angle=0,width=0.45\textwidth]{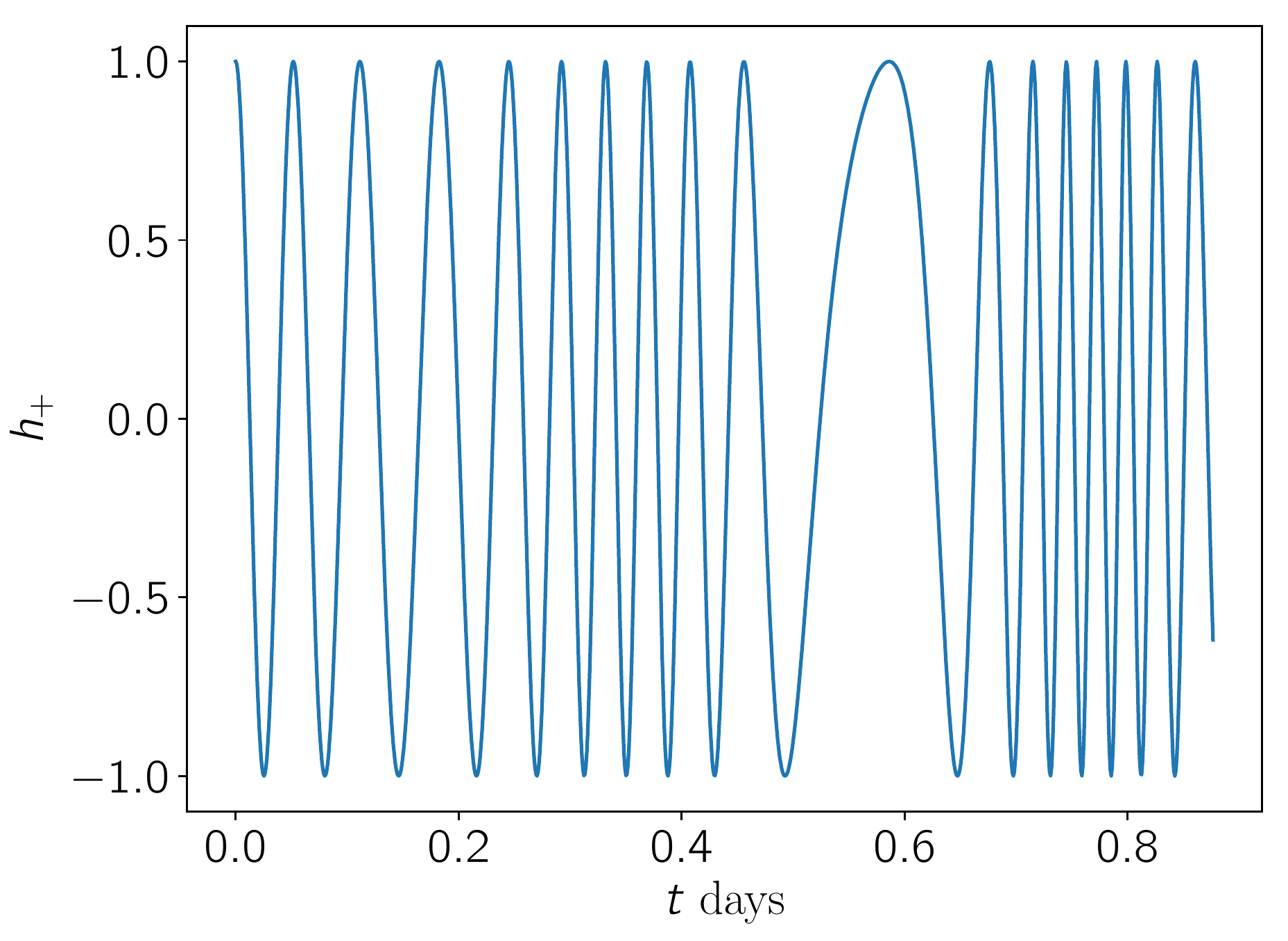} 
		\caption{\label{fig:triple_freq_evol} The gravitational waveform seen at the solar system barycenter for a system with outer period $P_{2}= 1.1$ hours and the line-of-sight velocity amplitude $0.1$. The carrier frequency $f$ of the gravitational wave in this example is $5$ mHz.}
	\end{centering}
\end{figure}

The presence of a perturbing companion star $m_{c}$ leads to an acceleration of the center-of-mass with respect to the barycenter frame, hence red-shifting the signal such that the gravitational wave phase gets modified

\begin{equation}
\Psi_{\mathrm{gw}} = 2\pi \int^{t} \left[1+ v_{\mbox{\tiny $\parallel$}}(t')\right]f_{\mathrm{gw}}(t')dt' + \phi_{0} \,\, ,
\label{eq:phase}
\end{equation}
where $v_{\mbox{\tiny $\parallel$}}$ is the line-of-sight velocity obtained in the previous section\footnote{Note that there should be an additional correction to Eq.~\eqref{eq:phase} due to the fact that time in the two frames is related by $t \mapsto t + r_{\mbox{\tiny $\parallel$}} / c$. This correction is however negligible for the triple systems considered in this work, but it might be relevant for systems closer to coalescence, e.g.~hierarchical triple black hole systems observable with LISA}. In Figure \ref{fig:triple_freq_evol} the quantity $h_{+}$ (normalized to $1$) is displayed for a circular triple system whose outer period was chosen to be very short to exaggerate the effects. The frequency oscillates around the carrier frequency $f$ modulating the gravitational wave phase. 

To accommodate these changes to the fast-slow waveform code~\cite{CornLitt07} must be modified. To properly calculate the gravitational wave transfer function one must evaluate it at the gravitational wave frequency observed by the LISA detectors. The Taylor expanded frequency evolution (as in equation (\ref{eq:f_src})) is redshifted with respect to the solar system barycenter i.e. $f_{\mathrm{gw}} \rightarrow \left(1+ v_{\mbox{\tiny $\parallel$}}\right)f_{\mathrm{gw}}$. The line-of-sight velocity is numerically obtained through equation (\ref{eq:v_LOS}) and the inversion of Kepler's equation. For the isolated galactic binaries the gravitational phase may be easily integrated. When this binary resides in a triple system an extra term in the gravitational wave phase integral crops up $2\pi \int v_{\mbox{\tiny $\parallel$}} f_{\mathrm{gw}}dt$ which is numerically integrated, interpolated at the detector sampling intervals, and then appended to the isolated galactic binary gravitational wave phase. These modifications to the gravitational wave frequency get applied to the slow portion of the waveform model, which is sampled at cadence much longer than the orbital period.

The log likelihood function used in our analysis involves noise-weighted inner products of the form
\begin{align}
(g|k) = 4\mathcal{R} \int_{0}^{\infty}\frac{\tilde{g}^{*}(f)\tilde{k}(f)}{S_{n}(f)} df \,\,,
\end{align}
where $g$ and $k$ are arbitrary waveforms as seen by LISA, and $S_{n}(f)$ is the one-sided noise power spectral density. Further discussion of this quantity and the noise model for LISA, including both instrumental noise and unresolved galactic binary confusion noise, can be found in references \cite{Cornish:2017vip, RobCorn17, CornRob18}. The signal-to-noise ratio (SNR) $\rho$ is defined as $\rho^{2} = (h|h)$ for a given waveform $h$.

\begin{figure}[htp]
	\begin{centering}
		\includegraphics[clip=true,angle=0,width=0.45\textwidth]{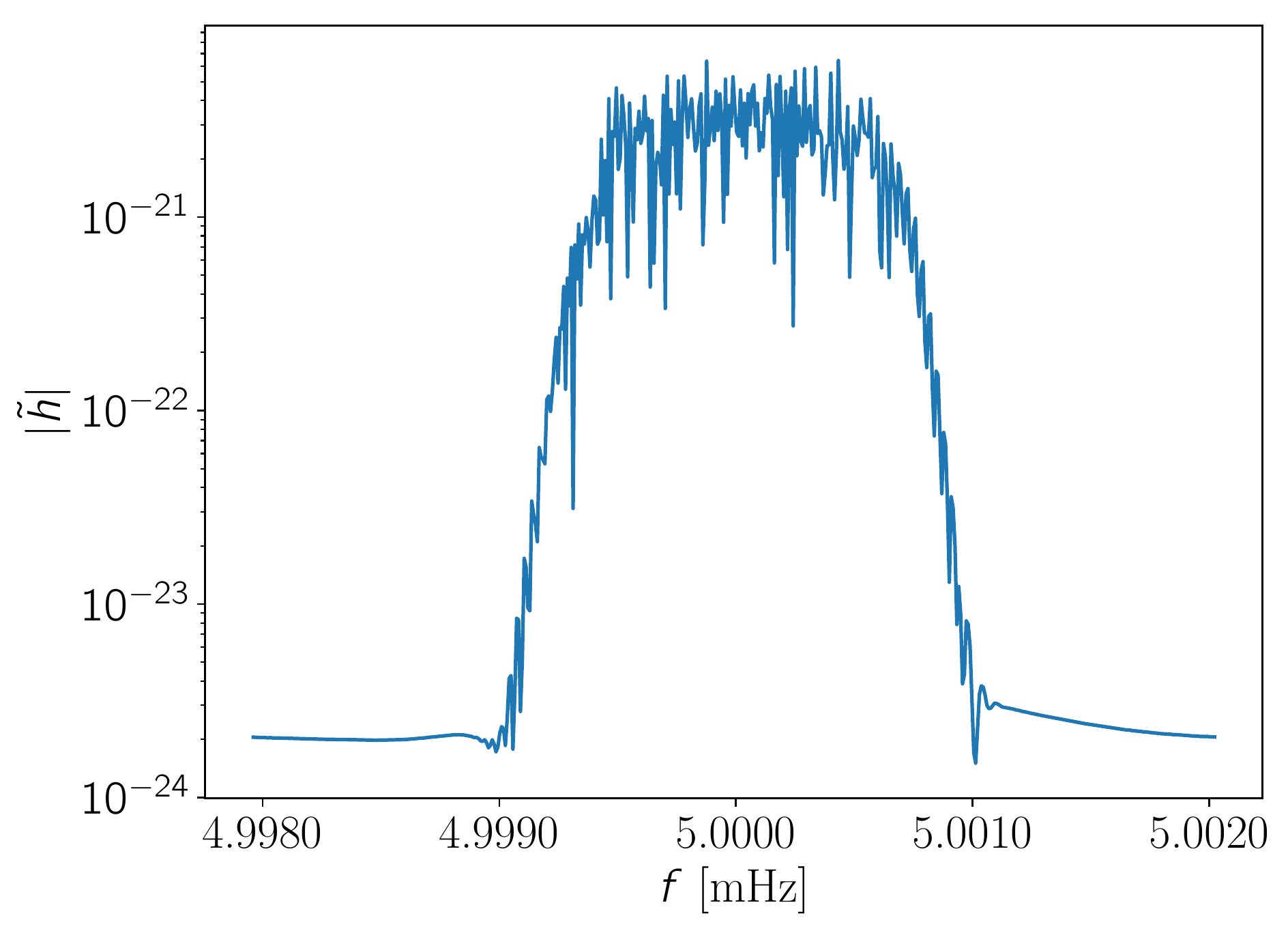} 
		\caption{\label{fig:triple_example1} Triple system with an outer orbital period $P_{2} = 1.5$ years and $e_{2} = 0.3$. The presence of the perturbing companion induces harmonics of the carrier frequency and of the harmonics present due to LISA's modulations.}
	\end{centering}
\end{figure}
\begin{figure}[htp]
	\begin{centering}
		\includegraphics[clip=true,angle=0,width=0.45\textwidth]{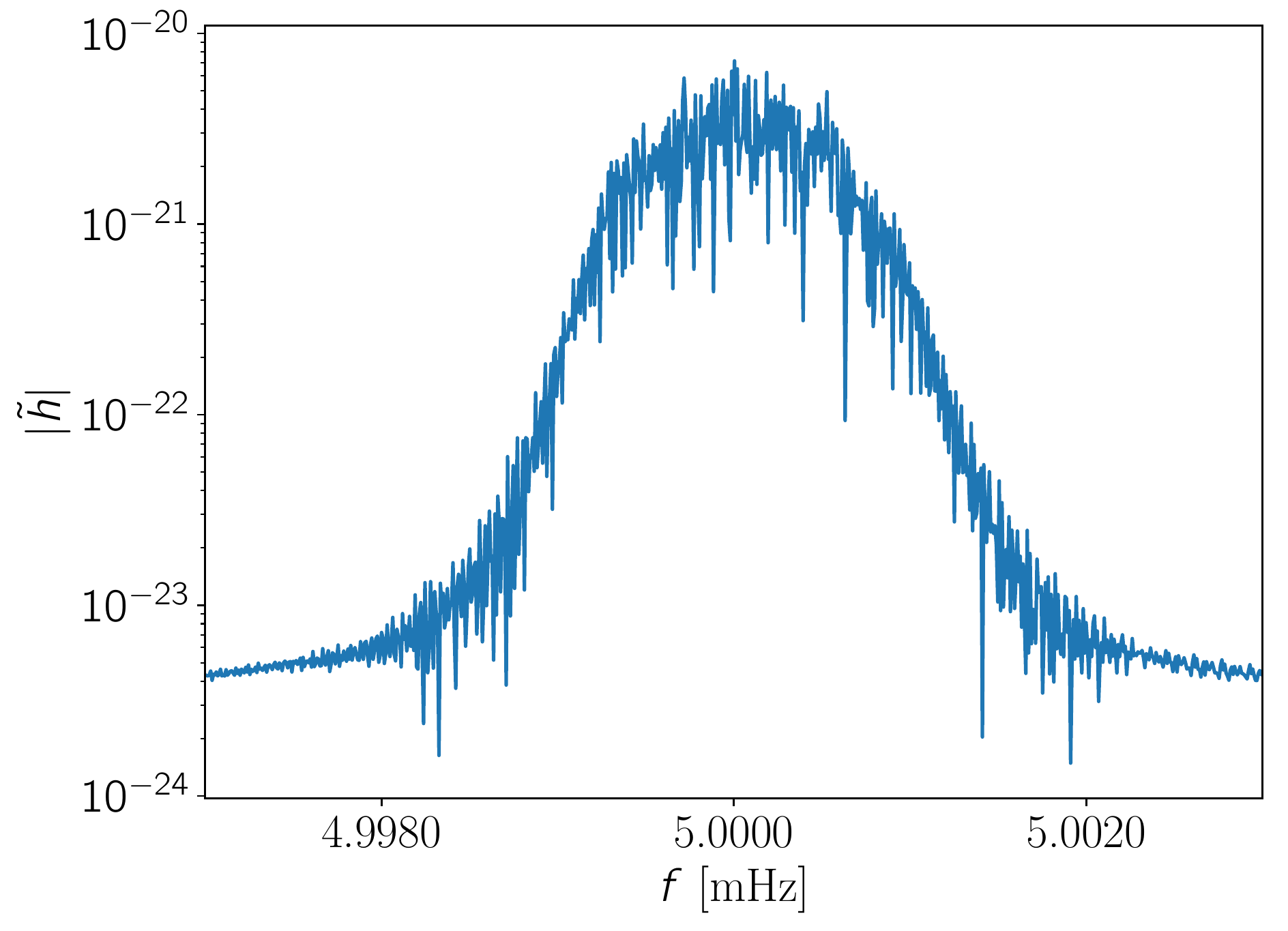} 
		\caption{\label{fig:triple_example2} Triple system with an outer orbital period $P_{2} = 0.6$ years and $e_{2} = 0.7$. In this example the harmonics induced by the companion star and LISA are interfering. Eccentricity in the outer orbit changes the distribution of power in the triple induced harmonics.}
	\end{centering}
\end{figure}

Examples of the frequency domain strain amplitude can be found in Figures \ref{fig:triple_example1} and \ref{fig:triple_example2}. Both of these waveforms were generated for inner binaries with $f = 5$ mHz, and a chirp mass of $0.32 M_{\odot}$ (which fixes the source frame frequency evolution as determined by General Relativity) for a $4$ year observation period at a $15$ second cadence. In Figure \ref{fig:triple_example1} the outer orbit revolves every $1.5$ years and has an eccentricity of $0.3$. An isolated binary is nearly monochromatic, resulting in a near delta function in the frequency domain, but due to the modulations of LISA as it cartwheels around the Sun in a year, picks up side-bands whose phase and relative amplitude are determined by the sky location and gravitational wave polarization of the binary. The introduction of a perturbing third body generates more harmonics of the frequencies already present, and tends to increase the bandwidth of the signal. Increasing the eccentricity of the outer orbit shifts the distribution of power into higher modes of the triple harmonics.

In Figure \ref{fig:triple_example2} the system has a tighter outer period of $0.6$ years, and a larger eccentricity of $0.7$. Here the orbital period $P_{2}$ is comparable to the orbital timescale of LISA, leading to a strong interference of harmonics. The side-bands of the carrier frequency induce by the triple are now more widely spaced than those imparted by LISA orbit, leading to a much broader signal, which is amplified by the larger eccentricity. If one were to consider a system of even a shorter period then the triple induced harmonics separate out into isolated side-bands.

\section{Detecting Hierarchical Companions}\label{detect}

When LISA first detects a triple the only the intrinsic frequency of the inner binary will be measurable. As more cycles are accumulated, and the center-of-mass of the inner binary has moved through a significant portion of the outer orbit, the data will support the inclusion of orbitally induced redshifts. We will now estimate when we expect the frequency evolution to be measurable, i.e.~for a given source and observation period, and an average oriented source, what $P_{2}$'s will we be able to detect with the effect of this center-of-mass motion?

From the gravitational wave phase quoted in equation (\ref{eq:phase}) it is straight-forward to obtain the frequency time derivative in the barycenter frame for a binary in a triple system which has no (or very little in comparison) source frame frequency evolution

\begin{equation}
\dot{f} = a_{1,\mbox{\tiny$\parallel$}} f \,\, ,
\label{eq:fdot_triple}
\end{equation}
where $a_{\mbox{\tiny$\parallel$}}$ is the line-of-sight acceleration of the inner binary's center-of-mass. This can obtained by differentiating equation (\ref{eq:v1_vector})

\begin{equation}
\textbf{a}_{1} = - \frac{m_{c}}{p_{2}^{2}}\left(1+e_{2}\cos\varphi_{2}\right)^{2}\left(\cos\varphi_{2},\sin\varphi_{2},0\right)\,\,,
\end{equation}
applying the rotation matrices once more and projecting along the line-of-sight gives

\begin{align}
a_{1, \mbox{\tiny$\parallel$}}  = -&\frac{m_{c}}{p_{2}^{2}}\left(1+e_{2}\cos\varphi_{2}\right)^{2} \nonumber\\
&\times \left[S\cos(\omega_{2}+\varphi_{2}) +C \sin(\omega_{2}+\varphi_{2})\right]\,\,,
\end{align}
where $S$ and $C$ are defined as before; Cf.~equation \eqref{eq:v_los}. We may square this quantity and then average it over the angles $\phi$, $\theta$, $\omega_{2}$, and $\iota_{2}$

\begin{align}
\left<a_{1, \mbox{\tiny$\parallel$}}\right>  =& \frac{1}{(4\pi)^{2}}\int_{0}^{2\pi}d\phi \int_{-1}^{1}d(\cos\theta) \nonumber \\
&~~\times  \int_{0}^{2\pi}d\omega_{2}\int_{-1}^{1}d(\cos\iota_{2}) a_{1, \mbox{\tiny$\parallel$}}^{2}  \\
=&\frac{m_{c}^{2}}{3 p_{2}^{4}}\left(1+e_{2}\cos\varphi_{2}\right)^{4}\,\,. 
\end{align} 
To calculate the RMS acceleration we average the previous result over the course of an orbit

\begin{align}
a_ {\mbox{\tiny $\parallel$ RMS}}^{2} &=\frac{1}{P_{2}}\int_{0}^{P_{2}} \left<a_{1, \mbox{\tiny$\parallel$}}\right> dt\\
&=\frac{1}{P_{2}}\int_{0}^{P_{2}} \left<a_{1, \mbox{\tiny$\parallel$}}\right>  \dot{\varphi}_{2}^{-1} d\varphi_{2} \\
&=\frac{m_{c}^{2}}{3 m_{2}^{4/3}} \left(\frac{2\pi}{P_{2}}\right)^{8/3} \frac{1+\frac{1}{2}e_{2}^{2}}{\left(1-e_{2}^{2}\right)^{5/2}} \,\,.
\end{align}

In the regime that $P_{2}>T_{\mathrm{obs}}$ we may begin to describe the gravitational wave frequency as modified by the binary residing in a triple by a Taylor expansion. Equation (\ref{eq:fdot_triple}), when averaged over angles and over an orbit, provides us with a rough estimate of the size of $\dot{f}$ for an average outer orbit orientation which started at an average spot in its orbit when LISA began to collect data. With this we can ascertain how many frequency bins this $\dot{f}$ estimate will evolve the carrier frequency through

\begin{align}
\dot{f}T_{\mbox{\tiny obs}}^{2} =&~573 \left( \frac{P_{2}}{1 \text{ yr}}\right)^{-4/3} \left(\frac{m_{c}}{1 M_{\odot}}\right) \left(\frac{m_{2}}{2 M_{\odot}}\right)^{-2/3} \nonumber\\
&\times \left(\frac{T_{\mbox{\tiny obs}}}{4 \text{ yr}} \right)^{2} \left(\frac{f}{5 \text{ mHz}}\right)\sqrt{\frac{1+\frac{1}{2}e_{2}^{2}}{(1-e_{2}^{2})^{5/2}}} \,\, .
\label{eq:fdot_trip}
\end{align}
In order to ascertain when this effect is measurable we utilize Fisher matrix estimates for the error in measurement of $\dot{f}$. The Fisher matrix, by the Cramer-Rao bound, provides an estimate of the covariance matrix (upon inversion of the Fisher matrix), thereby providing error estimates. The Fisher matrix is defined as

\begin{equation}
\boldsymbol{\Gamma}_{ij} = (h_{,i}|h_{,j})\,\,,
\end{equation}
where $h_{,i}$ are derivatives of the waveform with respect to parameter $\lambda^{i}$ and then evaluated at the true parameters. For a triple signal whose outer period is larger than LISA's observation period we may readily approximate the frequency evolution of the system by a Taylor expansion as we would for a mildly chirping isolated binary. This allows us to utilize the fast galactic binary waveform to calculate the Fisher matrix.

Seto \cite{Seto02} used a simple toy model for a Fisher matrix analysis to estimate the measurement errors in some of the galactic binary parameters. In appendix A we expand upon these results and investigate how the errors get inflated by including more parameters through the use of the full galactic binary model. We find that the $\dot{f}$ and $\ddot{f}$ errors become inflated through the inclusion of the full set of galactic binary parameters. The criterion which we use to determine whether $\dot{f}$ is a measurable parameter is that $\dot{f}$ must be larger than $3\sigma$ (as estimated by the Fisher matrix) compared to no frequency evolution at all. This results in the outer period being measurable when this is short enough as quoted in equation (\ref{eq:fdot_meas}) from the introduction. A fiduciary source with an outer orbital period of $40$ years would have a measurable frequency evolution by the time the nominal LISA mission concluded. When the outer eccentricity $e_{2} = 0.7$ the measurable outer periods become $P_{2} = 110$ years and less.

We may make similar applications of the Fisher analysis to ascertain when the gravitational wave carrier frequency becomes biased (i.e. differs from source frame value on average in a measurable way) for a given $P_{2}$. The RMS line-of-sight velocity is given by

\begin{equation}
v_{\mbox{\tiny $\parallel$ RMS}}^{2} =\frac{m_{c}^{2}}{3 m_{2}^{4/3}} \left(\frac{2\pi}{P_{2}}\right)^{2/3}  \,\, ,
\end{equation}
such that when

\begin{align}
P_{2} \lesssim 71.8& \text{ yrs}\left(\frac{\rho}{10} \cdot \frac{m_{c}}{1.0 M_{\odot}}\cdot \frac{f}{5 \text{ mHz}}\right)^{3} \nonumber \\
&\times\left(\frac{m_{2}}{2 M_{\odot}}\right)^{-2}\left(\frac{T_{\mathrm{obs}}}{4 \text{ yrs}}\right)^{3} \,\, , 
\end{align}
our measurements of the carrier frequency $f$ will be biased. This is potentially the most concerning result if one is interested in the orbital period distribution of the galactic binaries, as for sources which only have $f$ measured, this yields a quite large range of outer orbital periods which could bias the frequency measurement.

Another question of interest is when the parameter $\ddot{f}$ is measurable (recall that here we are only considering the frequency evolution coming from the center-of-mass motion). Upon measuring $f$, $\dot{f}$, and $\ddot{f}$ we have the best chance of determining the underlying physics for mildly evolving sources. The RMS jerk is given by

\begin{equation}
\dot{a}_{\mbox{\tiny $\parallel$ RMS}}^{2} =\frac{m_{c}^{2}}{3 m_{2}^{4/3}} \left(\frac{2\pi}{P_{2}}\right)^{14/3}  \frac{1+\frac{19}{2}e_{2}^{2}+\frac{69}{8}e_{2}^{4} + \frac{9}{16}e_{2}^{6}}{(1-e_{2}^{2})^{11/2}}\,\, .
\end{equation} 

which for fiduciary values becomes measurable when  

\begin{align}
P_{2} \lesssim 16.7 & \text{ yrs}\left(\frac{\rho}{10} \cdot \frac{m_{c}}{1.0 M_{\odot}}\cdot \frac{f}{5 \text{ mHz}}\right)^{3/7} \nonumber \\
&\times\left(\frac{m_{2}}{2 M_{\odot}}\right)^{-2/7}\left(\frac{T_{\mathrm{obs}}}{4 \text{ yrs}}\right)^{3/7} \,\, .
\end{align}

\begin{figure*}[!htb]
	\centering
	\includegraphics[width=0.45\textwidth]{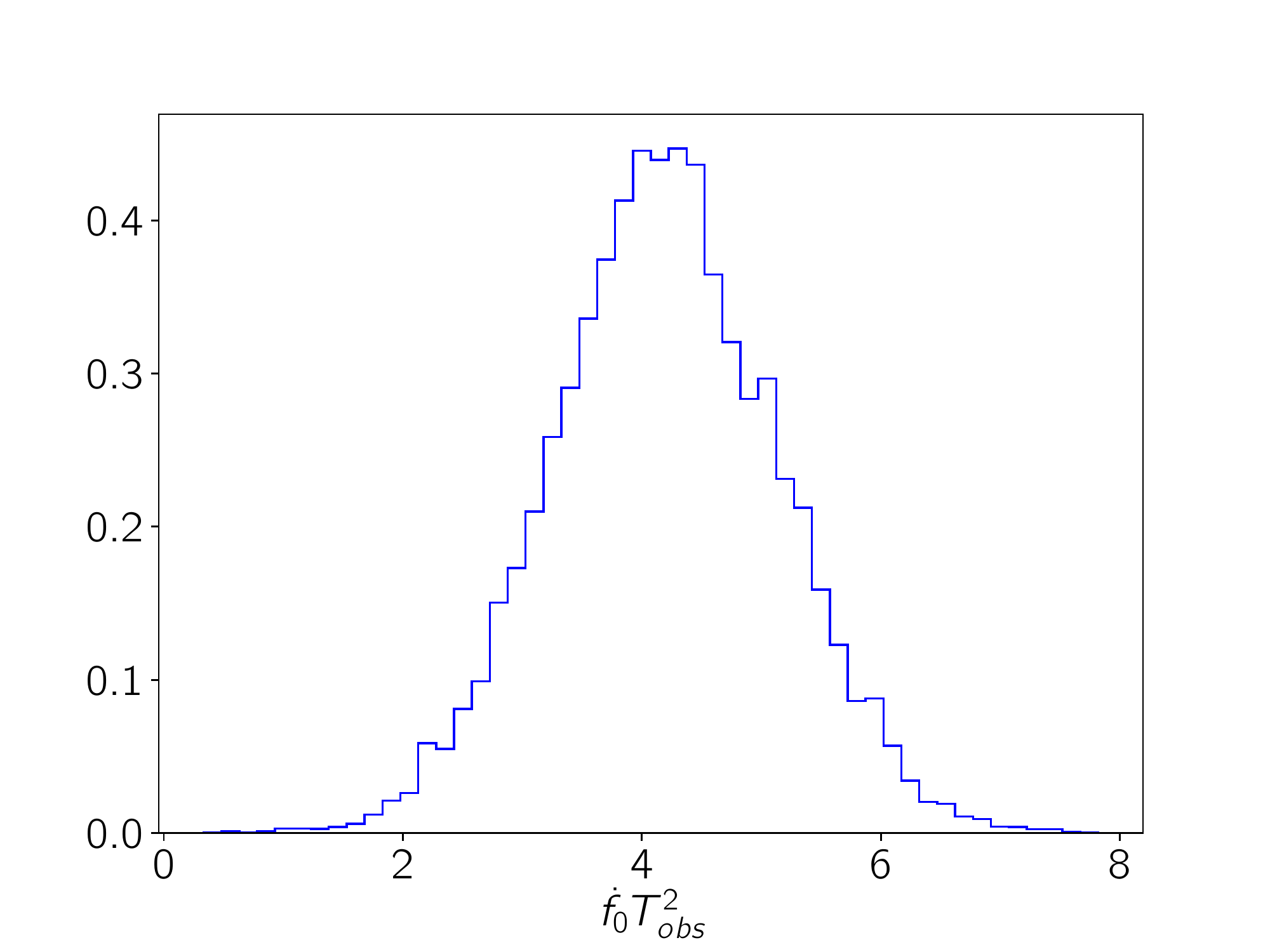} 
	\includegraphics[width=0.45\textwidth]{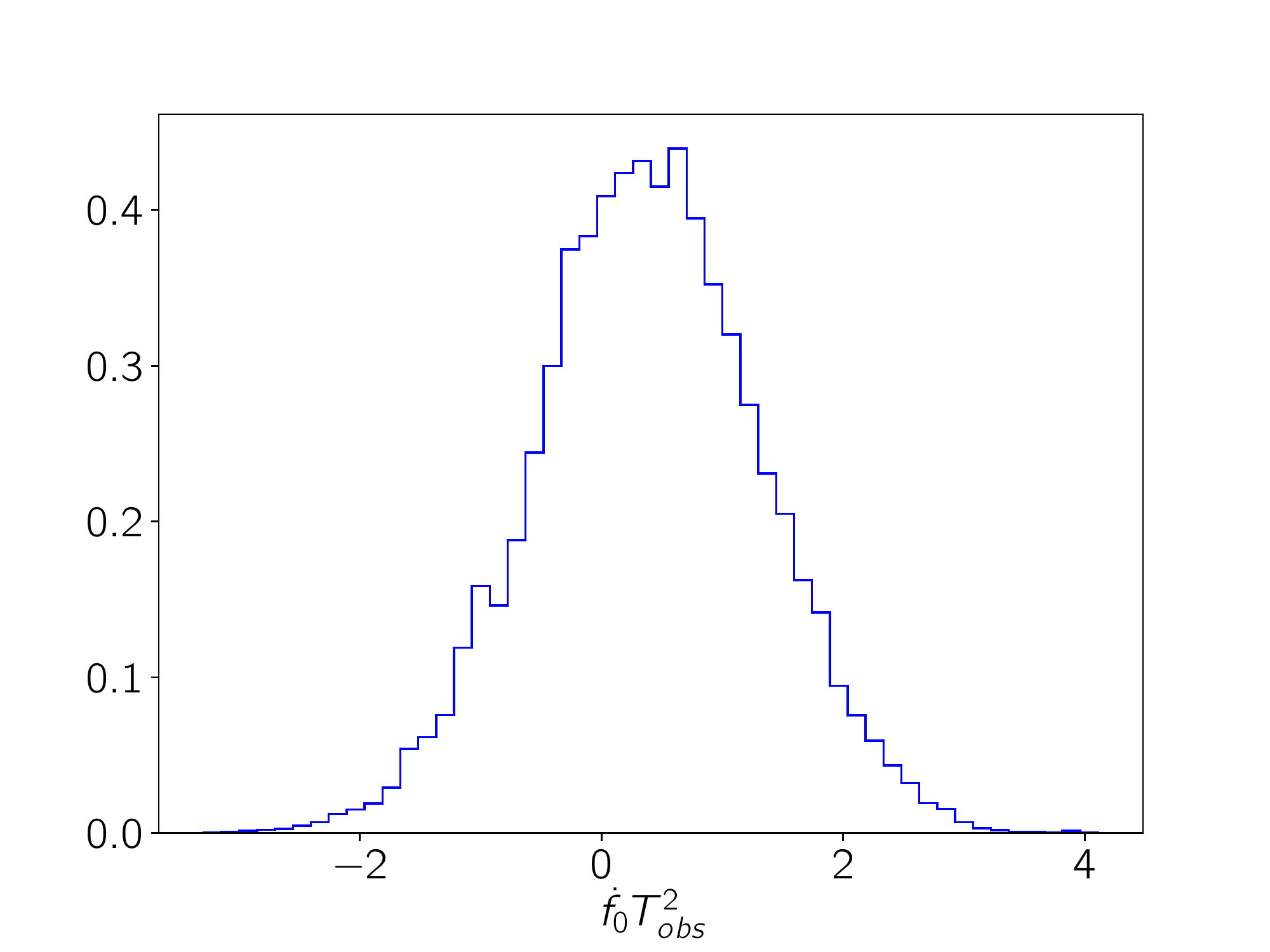} 
	\caption{\label{fig:fdot_measurable} These systems had a total mass of $m_{2} = 1.77 M_{\odot}$, with $\mathcal{M} = 0.32 M_{\odot}$, $\rho = 20$, and an observation period of $1$ year. The time of pericenter passage for the left hand figure was set to $0$ while for the right hand side $T_{2}$ was set to $-P_{2}/8$ i.e. an eighth of an orbit.}
\end{figure*}

To verify the validity of the preceding results, based off a Fisher matrix analysis, we now spot check the measurability of frequency evolution of a triple system using Markov Chain Monte Carlo (MCMC) simulations. Simulated data was produced for a triple system, and analyzed using the Taylor expanded frequency evolution model. The MCMC consisted of a burn-in phase such that the galactic binary model could search through parameter space to identify a regime in which the triple signal was well described by the binary model. A mixture of Fisher matrix proposal, differential evolution proposals and draws from the prior distribution were utilized to explore the posterior distribution~\cite{CornCrow15,KeyCorn11}. Parallel tempering was also used ito ensure a wide exploration of parameter space and to move between secondary modes of the posterior. 

In Figure \ref{fig:fdot_measurable} the posteriors for the parameter $\dot{f}$ (marginalized over all other parameters) are displayed for two triple systems. The outer period was chose to be $46$ years i.e. the value obtained from the fiduciary relation equation (\ref{eq:fdot_meas}) using the modified triple parameters. The errors predicted by the Fisher matrix for $\dot{f}$ are a bit smaller compared to the error measured by the MCMC, suggesting that we might be marginally overestimating the outer periods we can confidently measure. The difference between these two posteriors is the time of pericenter passage $T_{2}$ which differed by an eighth of an orbit between the two systems. This demonstrates that it is very important where we catch the triple in its orbit when LISA turns on, as the measurability of $\dot{f}$ is quite sensitive to $T_{2}$. This is especially important point to consider for larger outer period sources. Here we have seen that the Fisher analysis has roughly identified the regime in which we may hope to identify the presence of a triple system depending on where in the orbit we are measuring the gravitational wave signal.

\section{Characterizing the Hierarchical Orbit}\label{characterize}

Now that we have ascertained when the effects of a triple system are detectable we would like to know when the parameters of the triple orbit able to be measured. To determine this we utilized the Fisher information matrix for the triple signal. The criterion that we use to determine if a parameter is measurable is as follows: if the error in a parameter, as estimated by the Fisher matrix, is less than $50$\% of its true value then we claim this parameter can be measured. For triple systems the best measured parameter pertaining to the outer orbit is the outer orbital period, and if this quantity can be measured we say that the triple can be characterized (at least to some level). 

\begin{figure}[htp]
	\begin{centering}
		\includegraphics[clip=true,angle=0,width=0.5\textwidth]{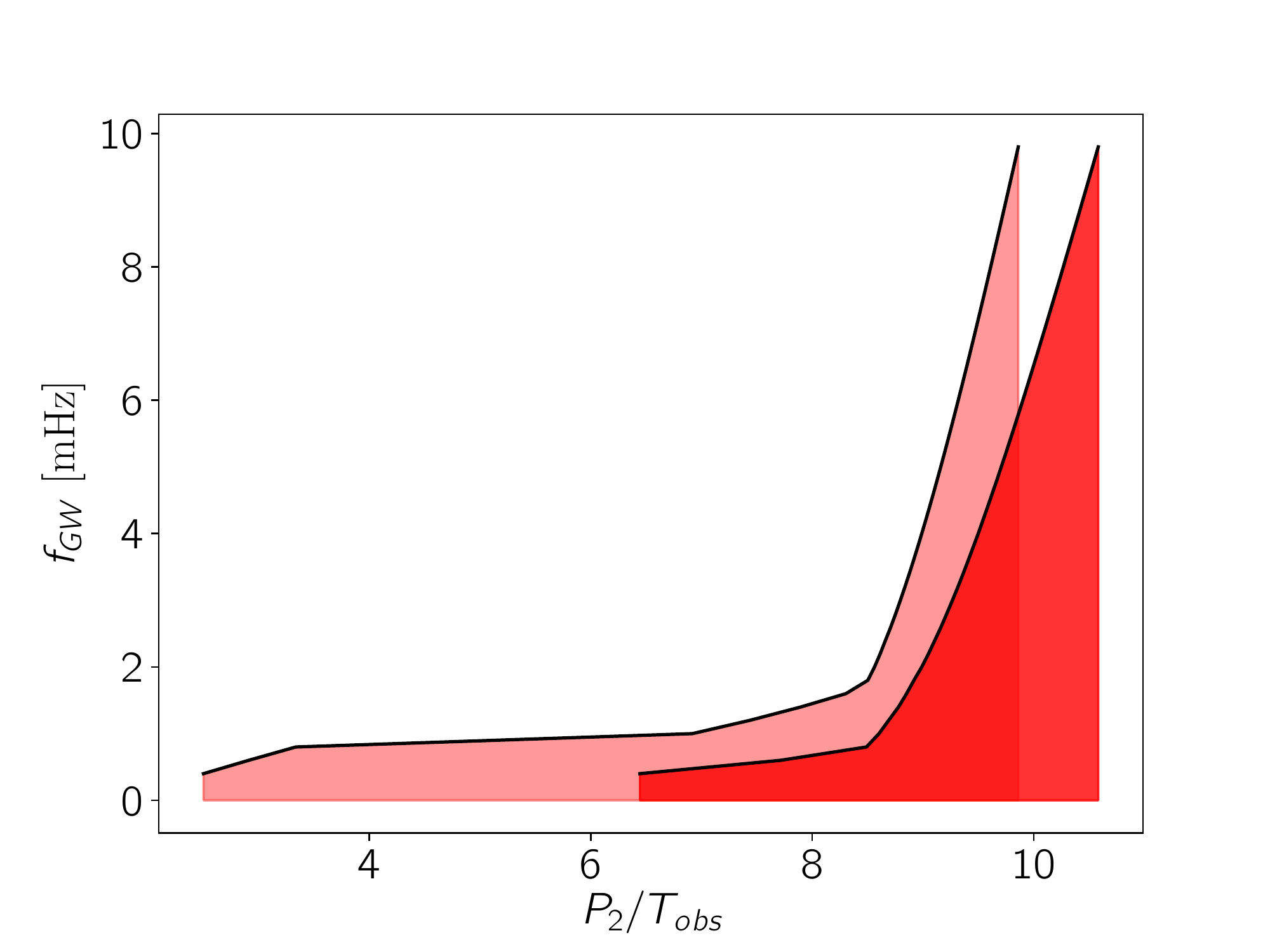} 
		\caption{\label{fig:char_triple} The leftward lines is for a SNR 20 system and the right line is for SNR 100. This system had the parameters  $m_{2} = 2.0 M_{\odot}$, $m_{c} = 1.0 M_{\odot}$, $m_{a} = 0.5 M_{\odot}$, and $\mathcal{M} = 0.32 M_{\odot}$}
	\end{centering}
\end{figure}

\begin{figure}[htp]
	\begin{centering}
		\includegraphics[clip=true,angle=0,width=0.5\textwidth]{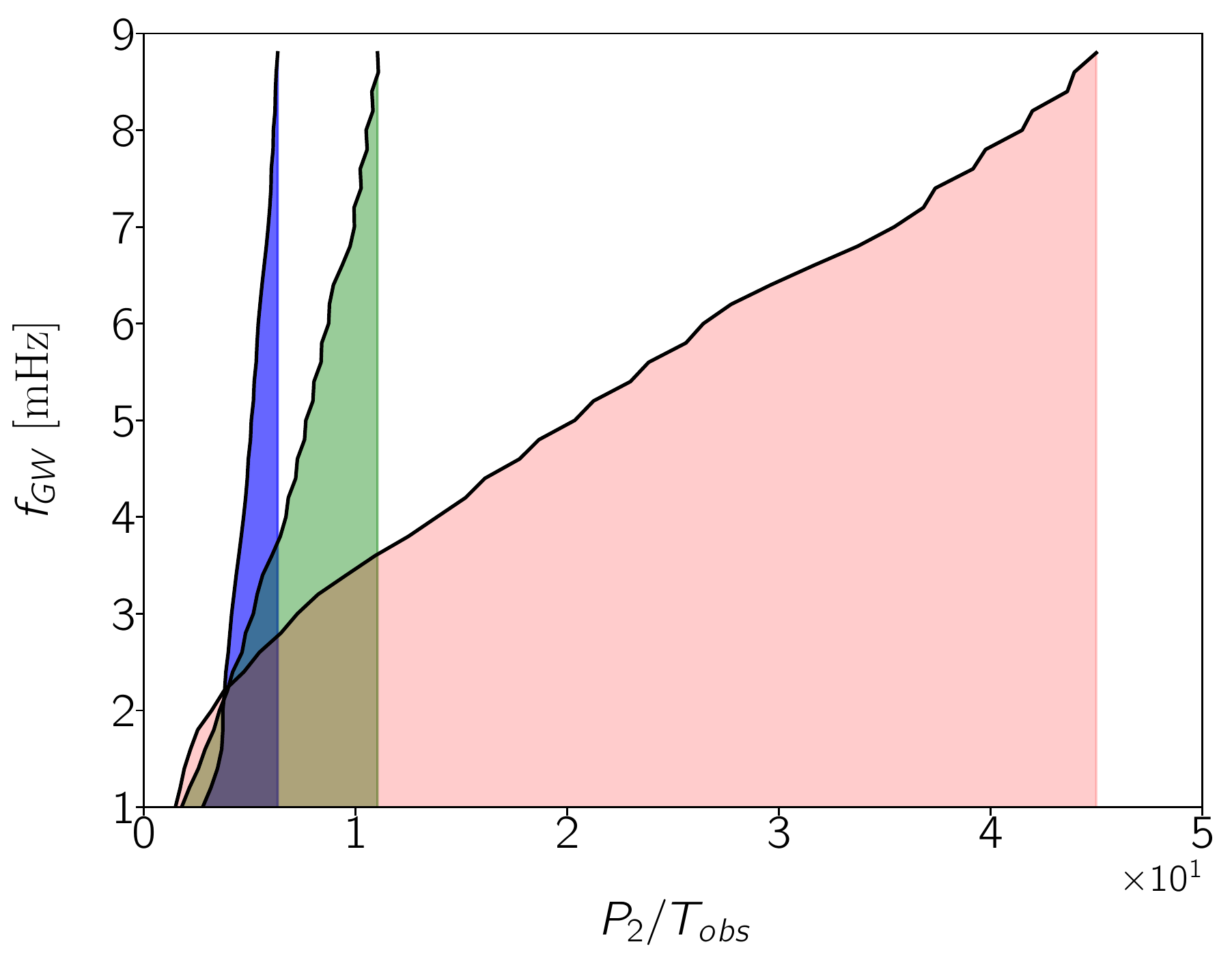} 
		\caption{\label{fig:vare2_triple_char} The red region denotes $e_{2} = 0.1$, blue $e_{2} = 0.4$, and green $e_{2} = 0.9$ These systems had an SNR of 10.
		}
	\end{centering}
\end{figure}

In Figure \ref{fig:char_triple} we display the results of the Fisher matrix based analysis. Systems with carrier frequencies and outer periods in the shaded region have orbits whose parameters \textit{cannot} be measured. To determine the separating line we construct a system with a given carrier frequency $f$ and a very short outer period $P_{2}$ and estimate its error with a Fisher analysis. The outer orbital is gradually made larger until its effects on the gravitational signal are marginal such that its error breaches $50$\%. The $P_{2}$ at which this happens defines the border in Figures \ref{fig:char_triple} and \ref{fig:vare2_triple_char}. We see that as the carrier frequency gets larger the outer period can be measured. This is due to this being a redshift phenomenon where the deviations in the frequency observed by LISA are proportional to the frequency itself, coupled with the fact that the error in the frequency is independent (to leading order, see appendix A) of the frequency itself.

Figure \ref{fig:vare2_triple_char} reveals the effect that eccentricity of the outer orbit has on the characterization of the triple parameters. Typically, for larger $f$, increasing the eccentricity allows one measure orbital periods that are larger than for the circular case. It is important to note that with such large orbits (in fact, any time when $P_{2} > T_{\mathrm{obs}}$) these results will depend on where we captured the triple in its orbits. For the systems considered here we chose $\varpi = 0$, and $T_{2} = 0$. This Fisher analysis demonstrates that we will be able to characterize the parameters for triple systems whose orbital period is up to $10$ times that of the LISA mission lifetime, though the details get slightly modified by the other parameters and SNR.

Let us now address how well the parameters of the triple system can be measured. The period and eccentricity of the outer orbit has the largest effect on the gravitational wave signal, and are therefore the most readily measured quantities. It is instructive to consider the strong parallels with the pulsar timing case. The analogy is clear; pulsars in a binary emit pulses at a very regular rate, with mild frequency evolution, and the arrival of these pulses gets modulated by Earth's motion and the presence of a companion. However, for pulsar timing the source is well localized on the sky, whereas the sky localization is generals poor for galactic binaries detected by LISA~\cite{Curt97}. Another parameter that is well measured in pulsar timing is ${(m_{\mathrm{b}}\sin\iota_{2})^{3}}/{m_{\mathrm{total}}^{2}}$, but it is only with the measurement of a Shapiro time delay for eclipsing binaries which allows the masses and inclination to be untangled. An additional affect, which will be negligible for the triples we are considering, is the variations in path length of light which allows the longitude of the ascending node $\Omega_{2}$ to be measured.

\begin{figure}[htp]
	\begin{centering}
		\includegraphics[clip=true,angle=0,width=0.5\textwidth]{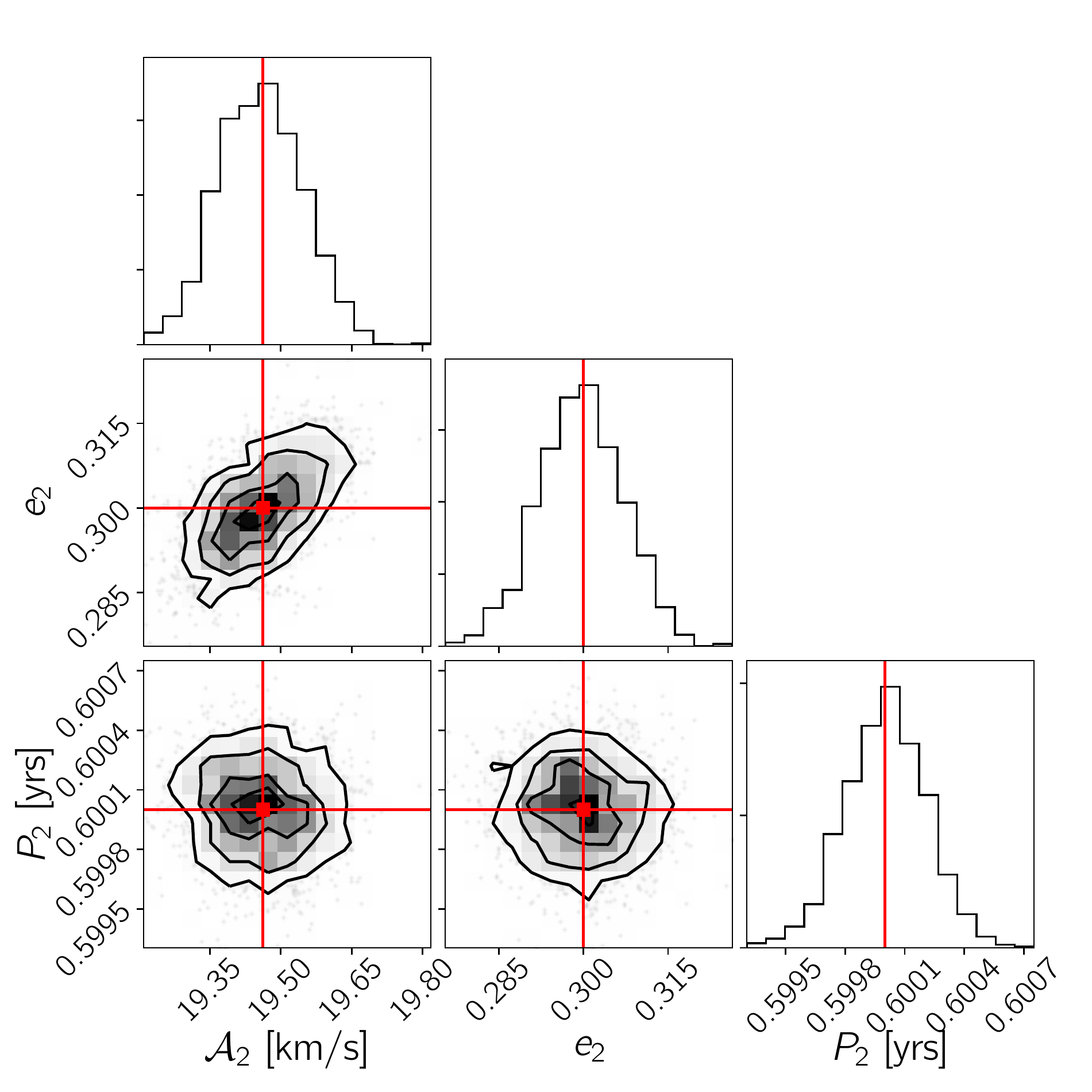} 
		\caption{\label{fig:triple_corner1} The line-of-sight velocity amplitude is $19.5$ km/sec, its eccentricity $0.3$, and outer period $0.6$ years. }
	\end{centering}
\end{figure}

\begin{figure}[htp!]
	\begin{centering}
		\includegraphics[clip=true,angle=0,width=0.5\textwidth]{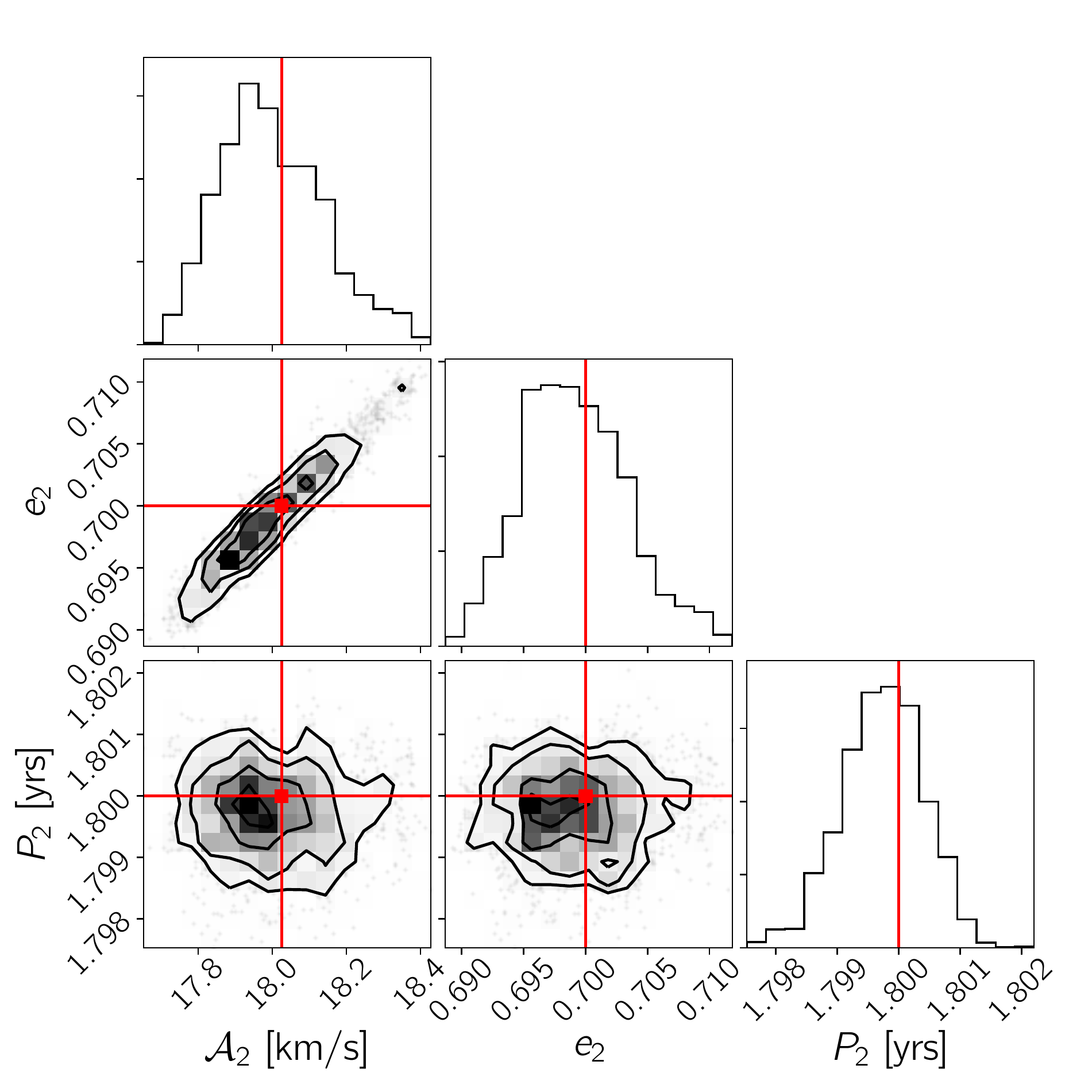} 
		\caption{\label{fig:triple_corner2} The line-of-sight velocity amplitude is $18.0$ km/sec, its eccentricity $0.7$, and outer period $1.8$ years. }
	\end{centering}
\end{figure}

\begin{figure}[htp]
	\begin{centering}
		\includegraphics[clip=true,angle=0,width=0.5\textwidth]{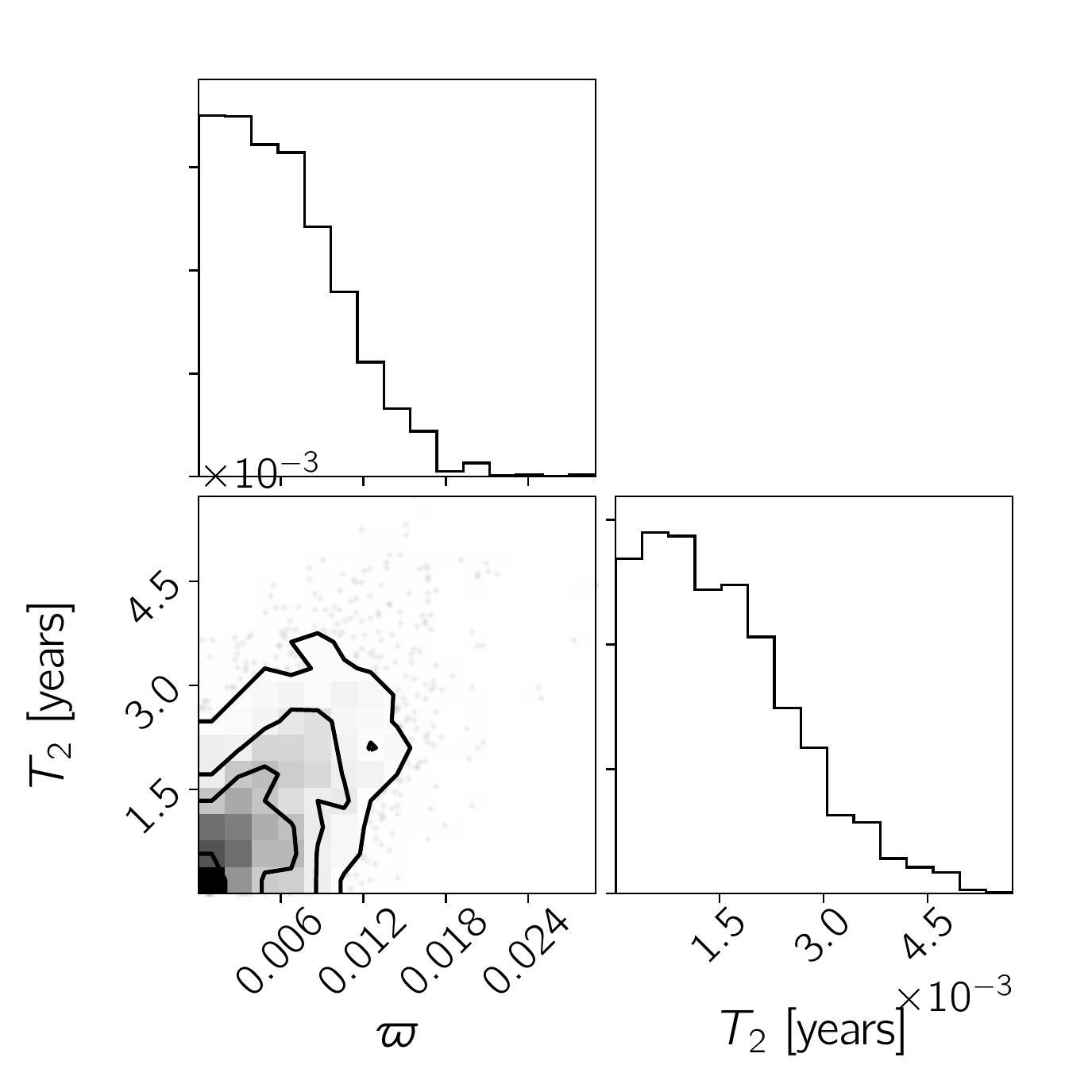} 
		\caption{\label{fig:triple_angles} Both $\varpi$ and $T_{2}$ were set to $0$ for this system. }
	\end{centering}
\end{figure}

Figures \ref{fig:triple_corner1} and \ref{fig:triple_corner2} are the results of MCMC of triple systems with an SNR of $50$ and a range of outer orbital periods and eccentricities. The injected values are marked by red lines and dots in these figures. As expected the outer period and eccentricity are well measured for both systems. For both of these systems the line-of-sight amplitude $\mathcal{A}_{2}$ is also well measured, but as discussed earlier, on its own not terribly informative. One sees that $\mathcal{A}_{2}$ and $e_{2}$ are correlated, which gets amplified in the more eccentric case. The fact that both of these parameters influence the amplitude of the harmonics induced by the triple is responsible for this correlation. In Figure \ref{fig:triple_angles} marginalized posteriors for the parameters $\varpi$ and $T_{2}$ are displayed for the more eccentric system. We see that these quantities are well measured, but they are of little physical interest.

One typically finds that as $P_{2}$ increases, such that less orbits are captured by LISA, the worse the parameters are characterized. In figure \ref{fig:triple_corner2} the eccentricity has a standard deviation of $0.6$\% relative to $e_{2}$ and the outer period an standard deviation of $0.037$\% or $14.6$ hours. The tighter system with $P_{2} = 0.6$ years had its outer orbital period determine to a standard deviation of $4.5$ hours. However, this does not seem to hold steadfast for the measurement of eccentricity. The tighter system had a relative standard deviation of $2.5$\% i.e. larger than the system with a wider orbit. This exception occurs as the outer orbital period starts to encroach upon the LISA modulation frequency (1 year). The distribution of power in the higher modes of the carrier frequency, induced by the triple, get shifted as $e_{2}$ changes. These harmonics, when their fundamental frequency $1/P_{2}$ is comparable to the LISA modulation frequency begin to interfere strongly making it harder to accurately extract the eccentricity.


\section{Ambiguous Systems}\label{ambiguous}

\begin{figure}[htp!]
	\begin{centering}
		\includegraphics[clip=true,angle=0,width=0.5\textwidth]{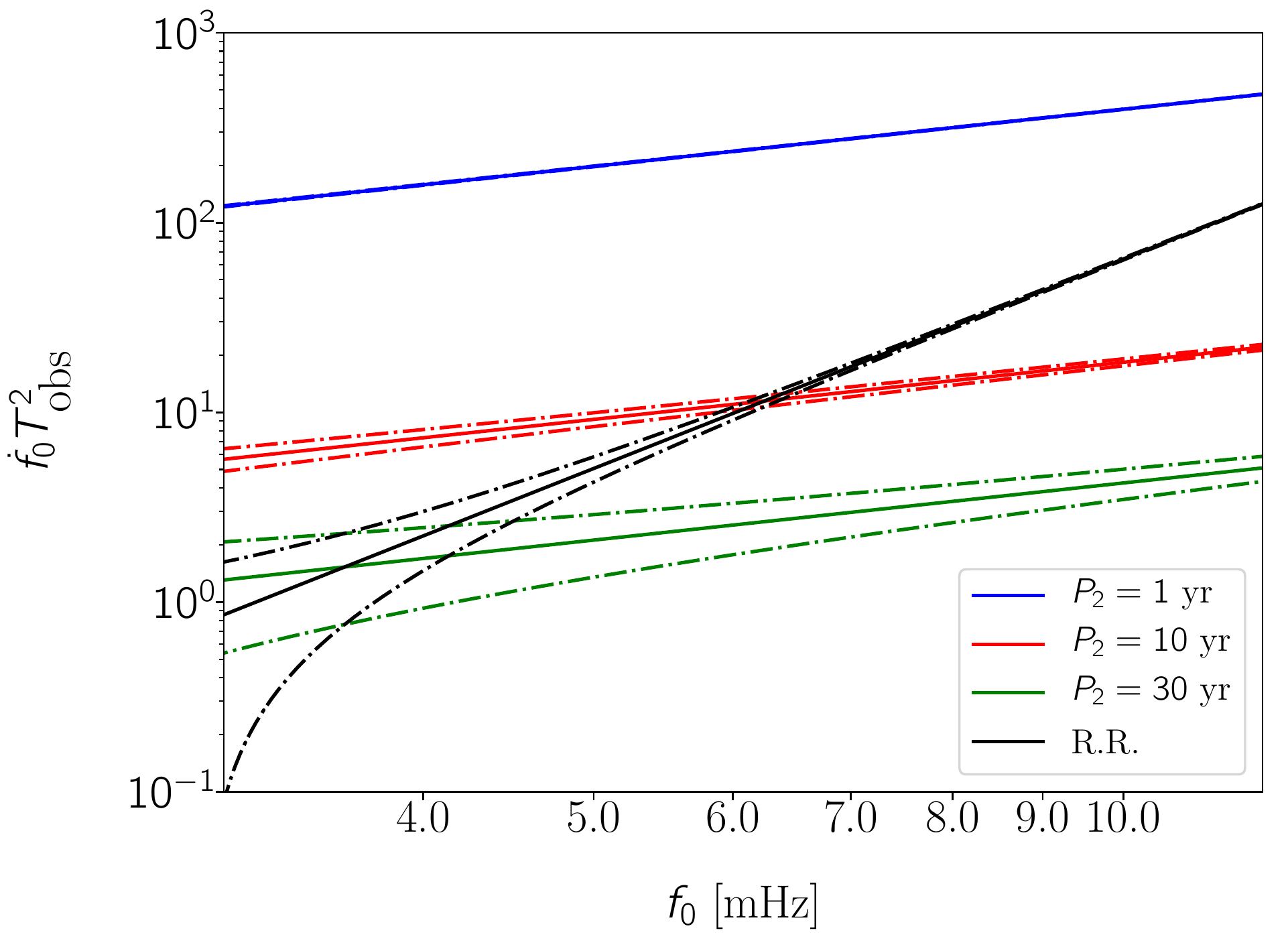} 
		\caption{\label{fig:f_fdot_track} The solid lines denote the gravitational wave frequency for an isolated binary and for several binaries in hierarchical orbits with outer periods of 1,10 and 30 years. In each case the outer orbit is circular, and the SNR of the gravitational wave signal is $\rho=50$. The dotted lines indicate the Fisher matrix error estimate for the frequency derivatives. Note the difference in power laws for the frequency derivatives and the small region of overlap between the curves.}
	\end{centering}
\end{figure}

Assuming a nominal $4$ year mission lifetime, its been estimated that frequency evolution due to gravitational wave emission or mass transfer will be measurable for roughly 9000 isolated galactic binaries~\cite{Cornish:2017vip}. It is interesting to consider if a regime exists where the orbital acceleration due to hierarchical companions may be confused with these effects.The chance of confusion is greatest with only $f$ and $\dot{f}$ are measurable. In most cases, a measurement of $\ddot{f}$ will break the degeneracy. To determine the risk of confusion consider Figure \ref{fig:f_fdot_track}, which compares the frequency evolution for an isolated binary and a binary in a hierarchical system. The frequency range over which the effects might be confused is very small since the frequency evolution scales very differently: $\dot{f} \propto  f$ from the hierarchical orbit (see equation (\ref{eq:fdot_triple})) and $\dot{f} \propto  f^{11/3}$ for mass transfer and gravitational wave emission.
We see that for an outer period of $1$ year there is no chance of confusion for this system. Even up to outer periods of $10$ years the amount of overlap is small. The system with an outer period of $30$ years, which is approaching the largest period for which there is a measurable $\dot{f}$, has the greatest potential for confusion. The larger the gravitational wave frequency the less likely it is that the effects will be confused. 

We now directly test how well a binary signal can reproduce a triple signal. To do so we inject a triple system into the LISA data stream and perform an MCMC with simulated annealing utilizing a galactic binary waveform model. The simulated annealing cools down the MCMC such that the chain settles into the peak of the posterior, thus allowing us to find the best values for the parameters as suggested by the data. The maximum posterior signal allows us to calculate the fitting factor

\begin{equation}
FF =\mathop{ \mathrm{max}}_{\boldsymbol{\lambda}} \frac{\left(h_{\mathrm{T}}|h(\boldsymbol{\lambda})\right)}{\sqrt{\left(h_{\mathrm{T}}|h_{\mathrm{T}}\right)\left(h(\boldsymbol{\lambda})|h(\boldsymbol{\lambda})\right)}} \,\,,
\end{equation}
where $\boldsymbol{\lambda}$ are the parameters which maximize the galactic binary model. The fitting factor is a measure of how well the maximum posterior galactic binary waveform $h(\boldsymbol{\lambda}_{\mathrm{max}})$ resembles the true triple waveform $h_{\mathrm{T}}$, which returns $1$ when the signals are equivalent and $0$ when they are perfectly orthogonal.

\begin{figure}[htp]
	\begin{centering}
		\includegraphics[clip=true,angle=0,width=0.5\textwidth]{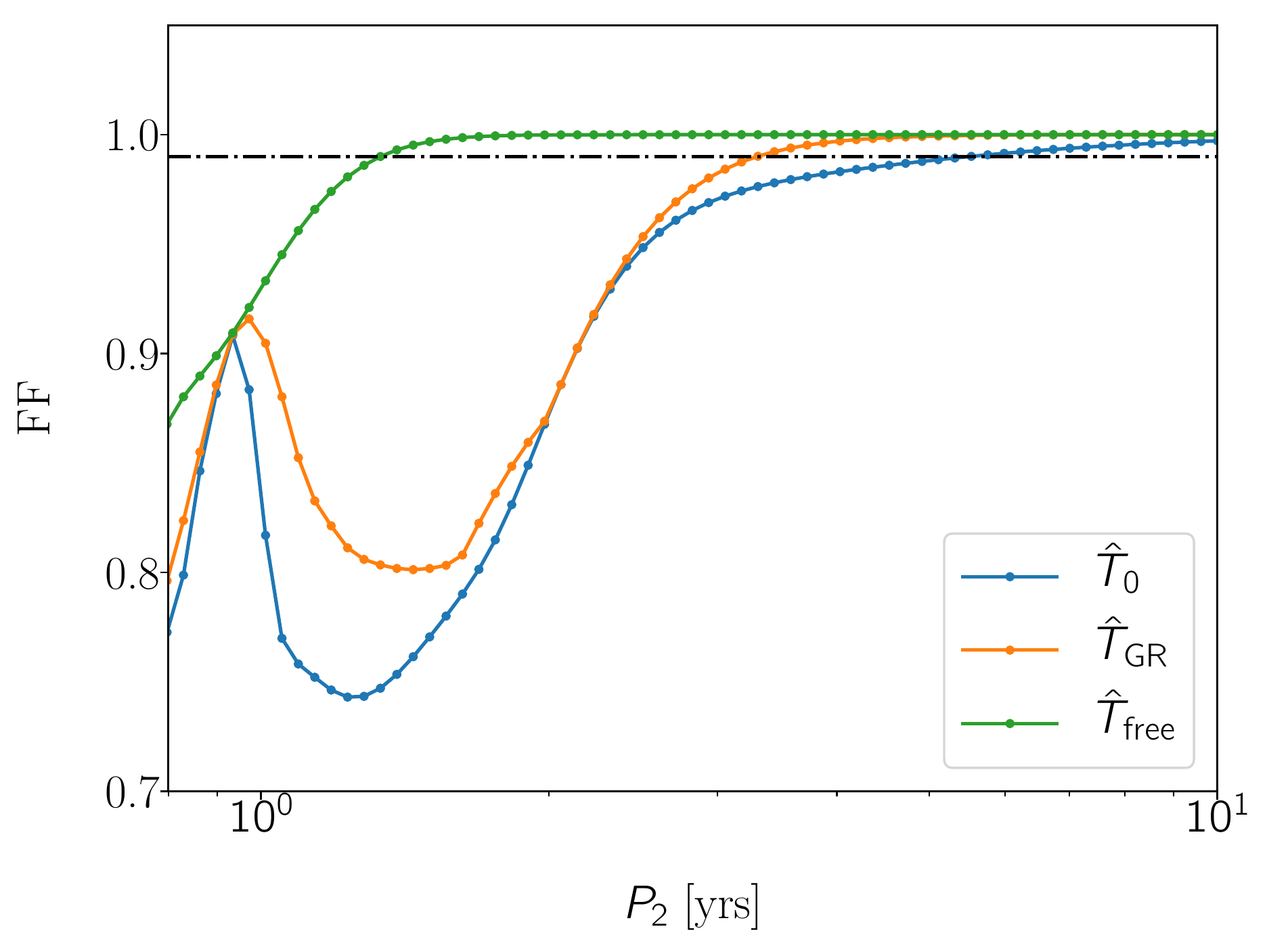} 
		\caption{\label{fig:FF_v_P2_1} The blue line represents the fitting factor for a purely monochromatic galactic binary model, the orange includes $\dot{f}$ in the frequency evolution, and the red includes $\dot{f}$ and $\ddot{f}$. These fitting factors are for circular systems.}
	\end{centering}
\end{figure}
In Figure \ref{fig:FF_v_P2_1} we show an example where the observation period was $1$ year and the carrier frequency $f$ was $3$ mHz for a circular outer orbit. The relevant masses for the triple were as follows: $m_{a} = 0.5 M_{\odot}$, $\mathcal{M} = 0.32 M_{\odot}$, $m_{c} = 1.0 M_{\odot}$. The parameters $T_{2}$ and $\omega_{2}$ were set to 0. There are three different models under consideration. The symbol $\hat{T}$ indicates models that uses a Taylor expanded frequency evolution
The $\hat{T}_{0}$ model assumes the signal is monochromatic i.e. it is characterized by only $f$.  The $\hat{T}_{\mathrm{GR}}$ model utilizes a three term Taylor expansion (i.e. $f$, $\dot{f}$, and $\ddot{f}$) in which the coefficients are related by the radiation reaction equations. Lastly, we consider the model $\hat{T}_{\mathrm{free}}$ which also utilized a three term Taylor expansion, but one in which there is no relation between the coefficients.

The $\hat{T}_{0}$ model is able to fit the signal from the hierarchical system for outer orbital periods that exceed $\sim4$ times the observation period, while the $\hat{T}_{\mathrm{GR}}$ model does a little better, and is able to fit the signal 
for outer orbital periods that exceed $\sim3$ times the observation period. The $\hat{T}_{\mathrm{free}}$ mode provides a good fit for outer orbital periods that exceed $\sim 1.2$ times the observation period. When the the outer period is comparable to, or short than the observation time the Taylor expansion representation of the frequency evolution will begin to fail, and we need to use the full orbital model. Note that in a time-evolving analysis of the LISA data, where the analysis is updated as the data arrives on Earth, the simple Taylor expansion model will initially work well for all systems, but as time goes on it will begin to break down for systems in hierarchical orbits. Long before that happens it will be obvious that these systems are part of a hierarchical system as the frequency derivatives will be far in excess of what we expect from mass transfer or gravitational wave emission (or equivalently, the chirp masses needed to explain the frequency evolution in terms of gravitational wave emission will be much larger than is expected for stellar remnants).

\begin{figure}[htp]
	\begin{centering}
		\includegraphics[clip=true,angle=0,width=0.5\textwidth]{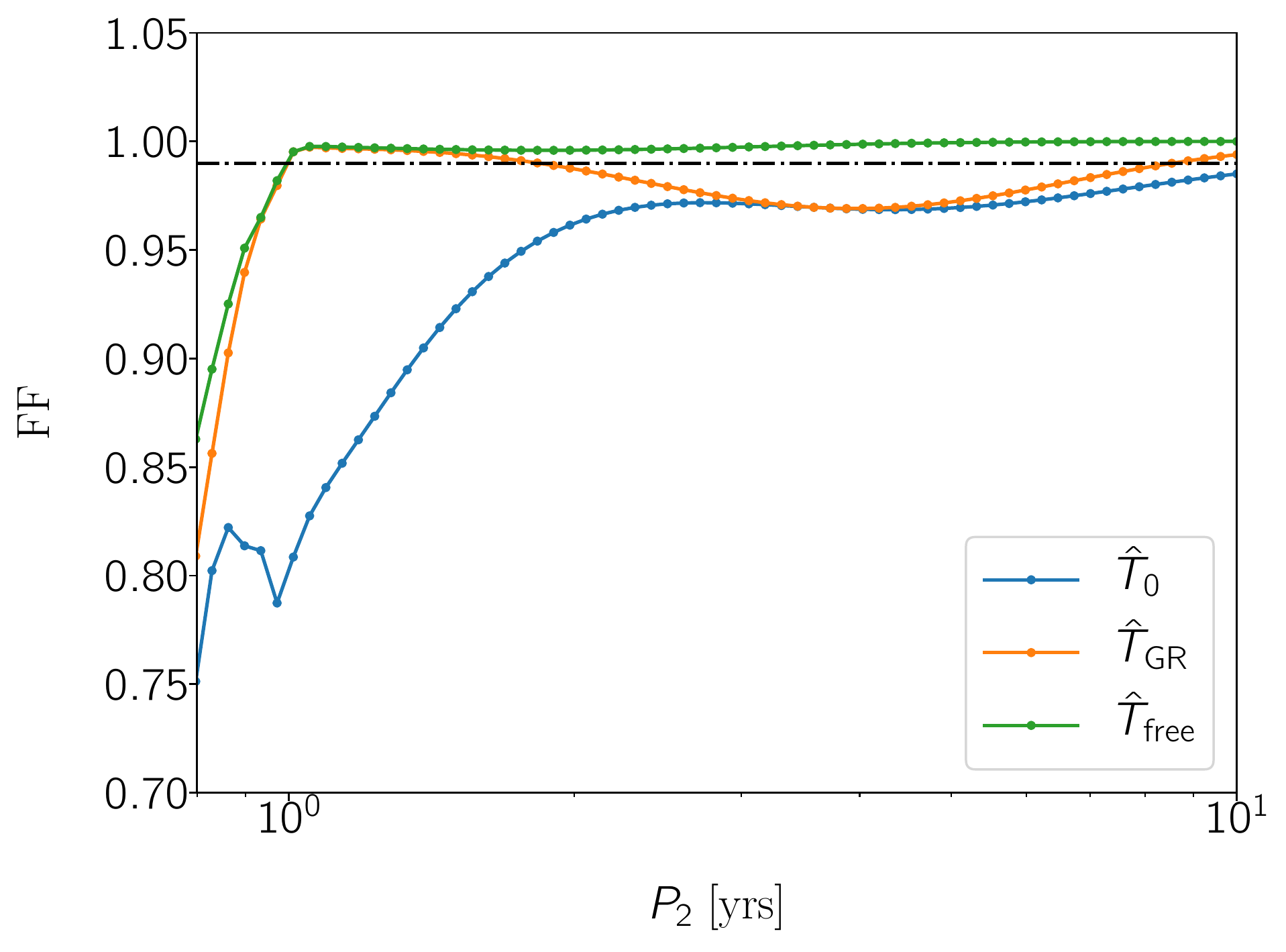} 
		\caption{\label{fig:FF_v_P2_2} These systems have a large eccentricity , $e_{2} = 0.7$. Fitting factors are larger compared to circular case.}
	\end{centering}
\end{figure}
The dash horizontal black line on Figure \ref{fig:FF_v_P2_1} denotes a fitting factor of $99\%$, which is what we expect for a perfectly modeled signal with SNR $20$. For a given SNR and model dimension $D$ (for which the galactic binary models we are considering vary from $7$ to $9$) the presence of noise will cause  the fitting factor to deviate from unity even with a perfect model for the signal. The expectation value for the fitting facto in the presence of noise is~\cite{Chatziioannou17}
\begin{equation}
{\rm FF} = 1-\frac{D-1}{2\rho^{2}} \,\, .
\end{equation}
Above the dashed line it  may not be possible to distinguish the Taylor expanded models from the full hierarchical model, though it will still be possible measure some parameters of the hierarchical orbit past where the dashed black line and fitting factor lines cross. Figure \ref{fig:f_fdot_track} allows one to see what outer periods could reproduce $\dot{f}$'s which resemble radiation reaction i.e. when the tracks overlap.

\begin{figure}[htp]
	\begin{centering}
		\includegraphics[clip=true,angle=0,width=0.5\textwidth]{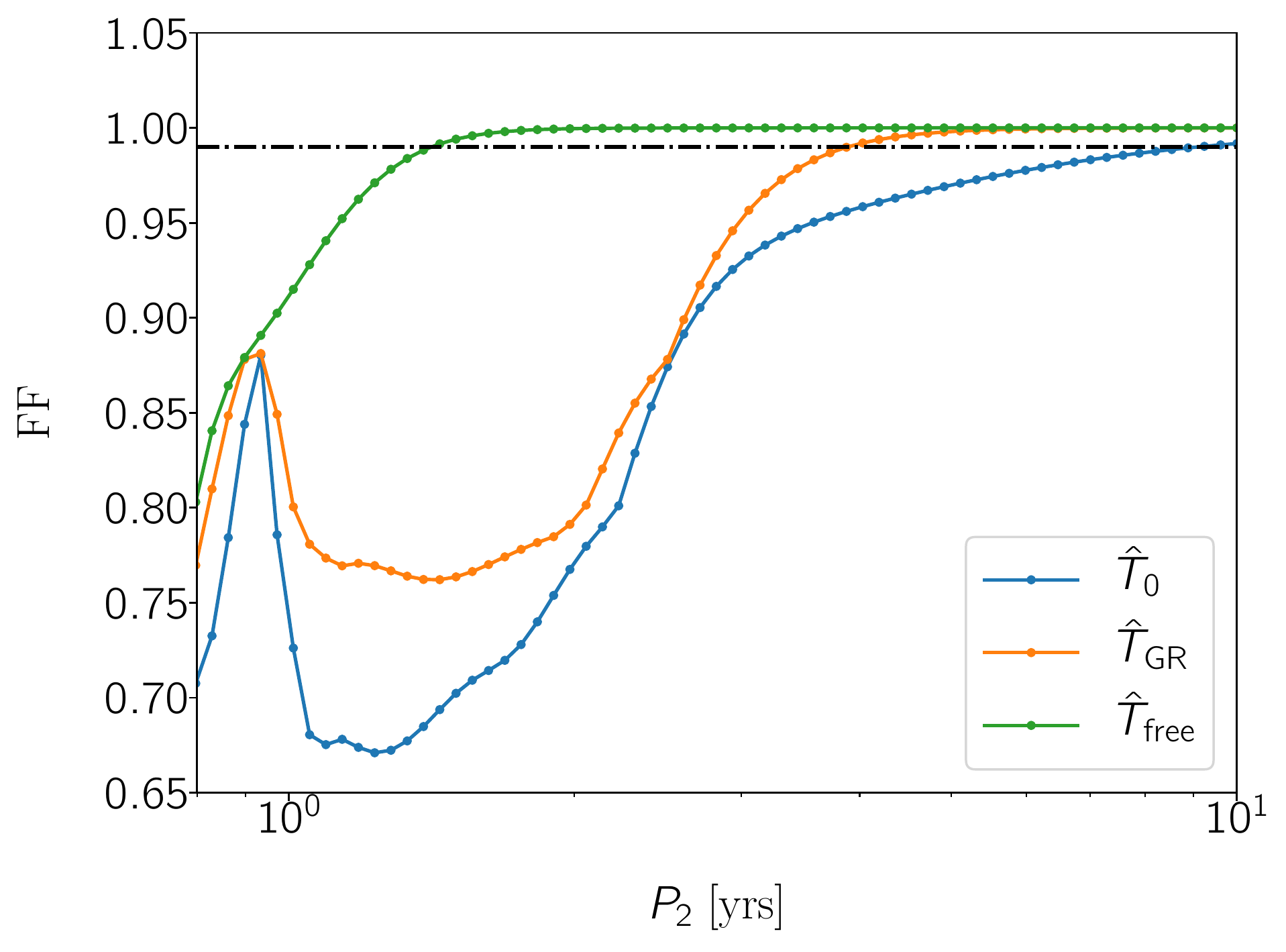} 
		\caption{\label{fig:FF_v_P2_3} These systems are circular, but the carrier frequency is $5$ mHz making the signal harder to replicate by a binary model.}
	\end{centering}
\end{figure}

In Figure \ref{fig:FF_v_P2_2} the eccentricity of the outer orbit is set to $0.7$. We see that the same general description holds. The fitting factors for the various Taylor expansion models decrease as $P_{2}$ approaches the LISA orbital timescale. The details of the interference's affects on the fitting factor change, and the fitting factors on the left side of the plot are generally a little higher. This is due to the shift of power to higher modes in the side-bands due to the larger eccentricity such that the most visible fundamental mode has less power. There is again no danger here of mis-modeling, as even larger frequency derivatives will be needed to accurately model these signals. Lastly, in figure \ref{fig:FF_v_P2_3} we see how a $5$ mHz source compares. We see that again the broad picture is intact, but the outer period at which the Taylor expanded models proves an ``acceptable'' fitting factor grows, leaving even less room for confusion between the models.

\section{Discussion}\label{discus}


Motivated by possibility that many of the galactic binaries observed by LISA may belong to hierarchical systems, we sought to answer three main questions (1) Under what circumstances can we detect the affects on a binary in a triple system (2) How well can we characterize the outer orbit of this system and (3) Understanding the regime of parameter space where we may confuse a triple system with an isolated binary. The frequency evolution incurred by a center-of-mass acceleration of the inner binary due to the presence of a perturbing companion will be measurable for outer periods as many as 10 times larger than the LISA mission lifetime. The outer orbital period and eccentricity will be measurable for systems whose outer periods are no larger than a few times the LISA mission lifetime. LISA will likely detect many triple systems and characterize their orbits, and in doing so provide unique insights into the role of that hierarchical companions have on binary evolution.

There will only be a small regime of parameter space in which we would expect to confuse the frequency evolution of an isolated binary with that imparted by a hierarchical companion. Analysis of the LISA data will require a global fit, simultaneously considering all detectable sources to account for covariances between the signals. One might be concerned about how the presence of binaries in hierarchical orbits will complicate the analysis, but it is only a mild complication. The simple Taylor expansion model will pick up the signals accurately at first, and once it becomes clear that the systems are undergoing large accelerations due to a distant companion, the signal model can be switch to the full orbital model.

There are many avenues for future research. For sufficiently tight systems, Kozai-Lidov oscillation or finite size effects may be measurable. Amother interesting scenario is that of eclipsing systems. In pulsar timing, an eclipsing system allows one to disentangle the mass and inclination of a binary through the measurement of time delays in the light. For a triple system, the eclipsing companion might induce a measurable Shapiro time-delay type effect into the gravitational wave, allowing us to learn more about the system.

\begin{appendices}
	\section{Basic Binary Fisher Analysis}

In this appendix we generalize the toy model introduced by Seto \cite{Seto02} which rather well approximates the errors in parameter estimation that one faces with a galactic binary signal in LISA. We model the signal as $h = A\cos\left(2\pi f_{\mbox{\tiny gw}} t + \phi\right)$, where $A$ is a constant amplitude, $\phi$ an arbitrary phase shift, and $f_{gw} = f + \dot{f} t +\frac{1}{2} \ddot{f} t^{2}$ (Note the difference in $\frac{1}{2}$ for the definition between our $\dot{f}$ and Seto's!). In this section we will investigate how the error analysis changes as we include more or less parameters in the model.

Under the assumption that the gravitational wave frequency is mildly chirping (such that the Taylor expansion is valid) you may approximate the noise-weighted inner product in the time domain as 

\begin{equation}
(g|k) = \frac{2}{S_{n}(f)}\int_{0}^{T}g(t)k(t)dt \,\,. 
\end{equation}
The Fisher matrix, in the approximation that many cycles are measured, i.e. $f T_{\mbox{\tiny obs}} \gg 1$, can be approximated as

\begin{equation}
\Gamma 
\approx \rho^{2}
\left(
\begin{matrix}
1 & 0 & 0 & 0 & 0  \\
0 & \frac{4}{3}\pi^{2} T^{2} & \pi^{2} T^{3} & \frac{2}{5}\pi^{2} T^{4} & \pi T \\
0 & \pi^{2} T^{3} & \frac{4}{5}\pi^{2} T^{4} & \frac{1}{3}\pi^{2}T^{5} & \frac{2}{3}\pi T^{2} \\
0 & \frac{2}{5}\pi^{2} T^{4} & \frac{1}{3}\pi^{2}T^{5}  & \frac{1}{7}\pi^{2}T^{6} & \frac{1}{4}\pi T^{3} \\
0 & \pi T & \frac{2}{3}\pi T^{2} & \frac{1}{4}\pi T^{3}  & 1 \\
\end{matrix}
\right) \,\, ,
\label{eq:fisher_full}
\end{equation}
where the matrix is ordered as $\log A, f, \dot{f},\ddot{f}, \phi$. Upon inversion we may obtain estimates of the errors in the parameters of interest by inverting the full Fisher matrix (or in versions where the $\ddot{f}$, and/or $\dot{f}$ dimensions are dropped). When only a monochromatic signal is used the RMS errors are

\begin{align}
\Delta f T_{\mbox{\tiny obs}} &= \frac{\sqrt{3}}{\pi} \rho^{-1} \approx 0.06 \left(\frac{10}{\rho}\right) \\
\Delta\phi &= 2\rho^{-1} \approx 0.20 \left(\frac{10}{\rho}\right)  \,\, .
\end{align}

Including $\dot{f}$ inflates the errors to the following

\begin{align}
\Delta f T_{\mbox{\tiny obs}} &= \frac{4\sqrt{3}}{\pi} \rho^{-1} \approx 0.22 \left(\frac{10}{\rho}\right) \\
\Delta \dot{f} T_{\mbox{\tiny obs}}^{2} &= \frac{3\sqrt{5}}{\pi} \rho^{-1} \approx 0.21 \left(\frac{10}{\rho}\right) \\
\Delta\phi &= 3\rho^{-1} \approx 0.30 \left(\frac{10}{\rho}\right)  \,\, .
\end{align}

Lastly, if one also includes the $\ddot{f}$ term

\begin{align}
\Delta f T_{\mbox{\tiny obs}} &= \frac{10\sqrt{3}}{\pi} \rho^{-1} \approx 0.55 \left(\frac{10}{\rho}\right) \\
\Delta \dot{f} T_{\mbox{\tiny obs}}^{2} &= \frac{18\sqrt{5}}{\pi} \rho^{-1} \approx 1.28 \left(\frac{10}{\rho}\right) \\
\Delta \ddot{f} T_{\mbox{\tiny obs}}^{3} &= \frac{20\sqrt{7}}{\pi} \rho^{-1} \approx 1.68 \left(\frac{10}{\rho}\right) \\
\Delta\phi &= 4\rho^{-1} \approx 0.40 \left(\frac{10}{\rho}\right)  \,\, .
\end{align}

Now we will consider a numerically calculated Fisher matrix for a galactic binary seen by LISA which includes only $f$, and $\dot{f}$ in its frequency evolution. The following matrix is ordered as $f$, $\cos\theta$, $\phi$, $\log A$, $\cos\iota_{1}$, $\psi$, $\phi_{0}$, and $\dot{f}$:

\begin{widetext}
	\begin{align}
	\Gamma = \left(
	\begin{matrix}
	~5.05 \times 10^{3} &~2.77\times 10^{2}   &-1.77\times 10^{2}  &~9.85\times 10^{-4} &~2.08                   &~2.49\times 10^{3} &~1.24\times 10^{3} &~1.82\times 10^{3}    \\
	~2.77\times 10^{2}   &~5.25\times 10^{2}  &-2.01\times 10^{2}  &-3.47                       &-3.56                    &~2.09\times 10^{1} &~1.04\times 10^{1} &~1.26\times 10^{2}  \\
	-1.77\times 10^{2}  &-2.01\times 10^{2}     &~3.15\times 10^{4} &~5.87\times 10^{-1}  &~1.37\times 10^{1} &-5.87\times 10^{1} &-2.88\times 10^{1} &~5.26\times 10^{2}\\
	~9.85\times 10^{-4} &-3.47                      &~5.87\times 10^{-1} &~4.00\times 10^{2}  &~4.13\times 10^{2} &-6.67\times 10^{-1} &~5.52\times 10^{-9} &~1.32\times 10^{-3} \\
	~2.08                      &-3.56                      &~1.37\times 10^{1}   &~4.13\times 10^{2}   &~4.27\times 10^{2}&~1.09 &~5.81e-01 & 1.04 \\
	~2.49\times 10^{3}   &~2.09\times 10^{1}  &-5.87\times 10^{1}  &-6.67\times 10^{-2}  &~1.09                     &~1.60\times 10^{3} &~8.00\times 10^{2} &~8.04\times 10^{2}\\
	~1.24\times 10^{3}   &~1.04\times 10^{1}   &-2.88\times 10^{1}  &~5.52\times 10^{-9} &~5.81\times 10^{-1} &~8.00\times 10^{2} &~4.00\times 10^{2} & ~4.02\times 10^{2} \\
	~1.82\times 10^{3}   &~1.26\times 10^{2}   &~5.26\times 10^{2} &~1.32\times 10^{-3} &~1.04                     &~8.04\times 10^{2} &~4.02\times 10^{2}&~6.95\times 10^{2}
	\end{matrix}
	\right) \,\, .
	\label{eq:fisher_binary}
	\end{align}
\end{widetext}
This system had a carrier frequency of $5$ mHz, a chirp mass of $0.32 M_{\odot}$, and SNR of $20$. The resulting Fisher matrix inverted, gives the following error estimates for the frequency

\begin{align}
\Delta f T_{\mathrm{obs}} &= 0.31 \,\,, \\
\Delta \dot{f} T_{\mathrm{obs}}^{2} &= 0.61 \,\,.
\end{align}
These errors are rather robust to choices in the parameters of the model. Comparing these results to the toy model considered above we see that the error in $f$ is roughly $3$ times larger when using the full galactic binary model and about $6$ times larger for $\dot{f}$. This results from the very strong covariance between $\phi_{0}$ and $\psi$ tied with the covariance of both of these parameters with $f$, and $\dot{f}$. If one considers galactic binaries modeled with $\ddot{f}$ as well one finds that the error in $\ddot{f}$ is about 4 times as great compared to the toy model estimate. These extra inflations are included in the analysis through the body of this paper in which we consider how tight the outer orbit must be for certain features to be measurable.

\end{appendices}

\section*{Acknowledgments}
TR and NJC appreciate the support of the NASA grant NNX16AB98G.
ST acknowledges support from the
Netherlands Research Council NWO (grant VENI [nr. 639.041.645]). We benefited from useful discussions with Valeria Korol.

\bibliography{refs}

\end{document}